\documentclass[final,5p,times,twocolumn]{elsarticle}
\usepackage{lineno,hyperref}
\usepackage{multirow}
\usepackage{multicol}
\usepackage{tikz}
 
\modulolinenumbers[5]

\journal{Journal of \LaTeX\ Templates}

\begin{document}

\begin{frontmatter}


\title{Developing Future Human-Centered Smart Cities: \\Critical Analysis of Smart City Security, Data Management, and Ethical Challenges}

\author[1]{Kashif Ahmad}
\author[2]{Majdi Maabreh}
\author[3]{Mohamed Ghaly}
\author[4]{Khalil Khan}
\author[5,6]{Junaid Qadir}
\author[1]{Ala Al-Fuqaha}
\address[1]{Information and Computing Technology (ICT) Division, College of Science and Engineering (CSE), Hamad Bin Khalifa University, Doha, Qatar}
\address[2]{Department of Computer Information Systems, Faculty of Prince Al-Hussein bin Abdullah II for Information Technology, The Hashemite University, Zarqa 13133, Jordan.}
\address[3]{Research Center for Islamic Legislation \& Ethics, College of Islamic Studies, Hamad Bin Khalifa University, Doha, Qatar.}
\address[4]{Department of Information Technology and Computer Science, Pak-Austria Fachhochschule: Institute of Applied Sciences and Technology, Haripur-KPK, Pakistan.}
\address[5]{Department of Computer Science and Engineering, Faculty of Engineering, Qatar University, Doha, Qatar.}
\address[6]{Department of Electrical Engineering, Information Technology University, Lahore, Pakistan.}

\ead{aalfuqaha@hbku.edu.qa}


\begin{abstract}

\textcolor{black}{As the globally increasing population drives rapid urbanisation in various parts of the world, there is a great need to deliberate on the future of the cities worth living. In particular, as modern smart cities embrace more and more data-driven artificial intelligence services, it is worth remembering that (1) technology can facilitate prosperity, wellbeing, urban livability, or social justice, but only when it has the right analog complements (such as well-thought out policies, mature institutions, responsible governance); and (2) the ultimate objective of these smart cities is to facilitate and enhance human welfare and social flourishing. Researchers have shown that various technological business models and features can in fact contribute to social problems such as extremism, polarization, misinformation, and Internet addiction. In the light of these observations, addressing the philosophical and ethical questions involved in ensuring the security, safety, and interpretability of such AI algorithms that will form the technological bedrock of future cities assumes paramount importance. Globally there are calls for technology to be made more humane and human-centered.  In this paper, we analyze and explore key challenges including security, robustness, interpretability, and ethical (data and algorithmic) challenges to a successful deployment of AI in human-centric applications, with a particular emphasis on the convergence of these concepts/challenges. We provide a detailed review of existing literature on these key challenges and analyze how one of these challenges may lead to others or help in solving other challenges. The paper also advises on the current limitations, pitfalls, and future directions of research in these domains, and how it can fill the current gaps and lead to better solutions. We believe such rigorous analysis will provide a baseline for future research in the domain.}

\end{abstract}


\end{frontmatter}


\section{Introduction}
\label{sec:introduction}
According to a recent report \cite{urbanization}, around 54\% of the world's population lives in cities, and the number is expected to reach 66\% by 2050. Rapid urbanization is driven by economic incentives but it also has a significant collateral environmental and social impact. Therefore, environmental, social, and economic sustainability is very crucial to maintain a balance between rapid expansion in the urbanization and resources of the cities. Thanks to modern technologies, striving for an improvement in the environmental, financial, and social aspects of urban life, and mitigating the associated challenges. More recently, the concept of smart cities has been introduced, which aims to make use of modern technologies including a wide range of Internet of things (IoT) sensors to collect and analyze data on different aspects of urban life \cite{gharaibeh2017smart,green2019smart}. A smart city application demands a joint effort of people from different disciplines, such as engineering, architecture, urban design, and economics, to plan, design, implement, and deploy a smart solution for an underlying task.

Artificial Intelligence (AI) techniques have also been proved very effective to gain insights from data collected through different IoTs sensors to manage and utilize the resources more efficiently. \textcolor{black}{In this paper, we use the term AI broadly as an umbrella term including techniques and algorithms able to learn from data (i.e., data science, statistical learning, machine learning, deep learning) or intelligent systems able to perform tasks such as perception, reasoning, inference (i.e., expert systems, probabilistic graphical models, Bayesian networks).} According to Greg Stone \cite{ARUP}, ``If you know the right questions and understand the risks, data can help build better cities,'' and AI helps you extract such insights from the data. Some key smart city applications where AI has been proved very effective include healthcare, transportation, education, environment, agriculture, defense, and public services \cite{qayyum2020secure,veres2019deep,xie2019survey,ahmad2020artificial,ullah2020applications}. We note here that while our focus in this paper is on AI, our ideas apply more broadly to the the more general case of artificial intelligence (AI) technology for smart cities in general. We also note that most of the recent significant advances have been made possible using advances in AI. Much of the work on AI safety and AI ethics is directly relevant to AI: therefore, we will mostly use both of these term synonymously. In a smart city application, AI techniques aim to process and identify a pattern in data obtained from individual sensors or collective data generated by several sensors, and provide useful insights on how to optimize underlying services. For instance, in transportation, AI could be used to analyze data collected from different parts of a city (e.g., roads, commute mode, and number of passengers) for future planning or deploying different transportation schemes in the city. 

However, there are several risks and challenges, such as availability, biases, and privacy of data, to successfully deploy AI in different smart city applications \cite{latif2019caveat,ekbia2015big,crawford2016there}. Various data biases can result in detrimental AI predictions in sensitive human-centric applications---e.g., algorithmic predictions may be biased against certain races and genders as reported in \cite{machine_bias} \cite{crawford2016artificial}. Apart from data-oriented challenges, there are some other threats to AI in smart cities' applications. For instance, attackers can launch different types of adversarial attacks on AI models to affect their predictive capabilities. Such attacks in sensitive application domains such as connected autonomous vehicles can lead to significant loss in terms of human lives and infrastructure \cite{qayyum2020securingCAV}. 

Another key challenge to the deployment of AI in smart city applications is the lack of interpretability (i.e., humans are unable to understand the cause of an AI model's decision) \cite{yang2017explainable}. Explainability is a key characteristic of AI models to be deployed in critical smart city applications, where the predictive capabilities of the models are not enough to solve a problem completely rather reasons behind the prediction are needed to be understood \cite{lundberg2018explainable,selvaraju2017grad}. It also helps to ensure that the AI decisions in an underlying application are equitable by avoiding decisions based on protected attributes (e.g., race, gender, and age, for instance, Amazon's AI recruiting tool was found biased against women \cite{amazon_bias}; similarly Amazon's Face Rekognition, a gender recognition tool, was found 31.4\% less accurate in classifying the gender of dark-skinned women compared to light-skinned men \cite{Face_Recognition,buolamwini2018gender}), and ensuring an equal representation of protected attributes in the sample space \cite{corbett2018measure}. In recent years, ever-growing concerns have been noticed on the deployment of AI algorithms in human-centric smart city applications, for instance, to ensure privacy issues in surveillance systems, unequal inclusion of citizens in different services, and biases in predictive policing \cite{kitchin2016ethics,o2016weapons,Face_Recognition}. 

\textcolor{black}{A breakthrough in one of these challenges may have a knock-on effect. For instance, to offset the problem of the interpretability and biases in decisions, explainable AI may also help to guard against adversarial attacks, and the explanations produced by explainable AI, on the other hand, may also help the attackers to generate more adverse attacks \cite{ignatiev2019relating,fidel2019explainability}.} 

\subsection{AI-Based Smart City Applications}
\label{sec:smart_cities}
In a smart city, sensors are deployed at various places to gather data about different aspects of the city---e.g., data related to transportation, healthcare, and environment---which is then sent to a central server for analysis or processed locally at the edge devices to obtain useful insights using AI techniques. Thanks to the recent advancement in technology, government authorities can now gather real-time data, combined with the capabilities of AI, can manage public services in cities more efficiently and effectively. For instance, having enough information about roads condition, traffic volume, and people's commute means in a city, authorities can eliminate the bottlenecks which can, in turn, reduce city traffic, crowd, and pollution, leading to a more optimized, sustainable, and clean services and environment. 

\begin{figure}[!h]
\centering
\includegraphics[width=.99\linewidth]{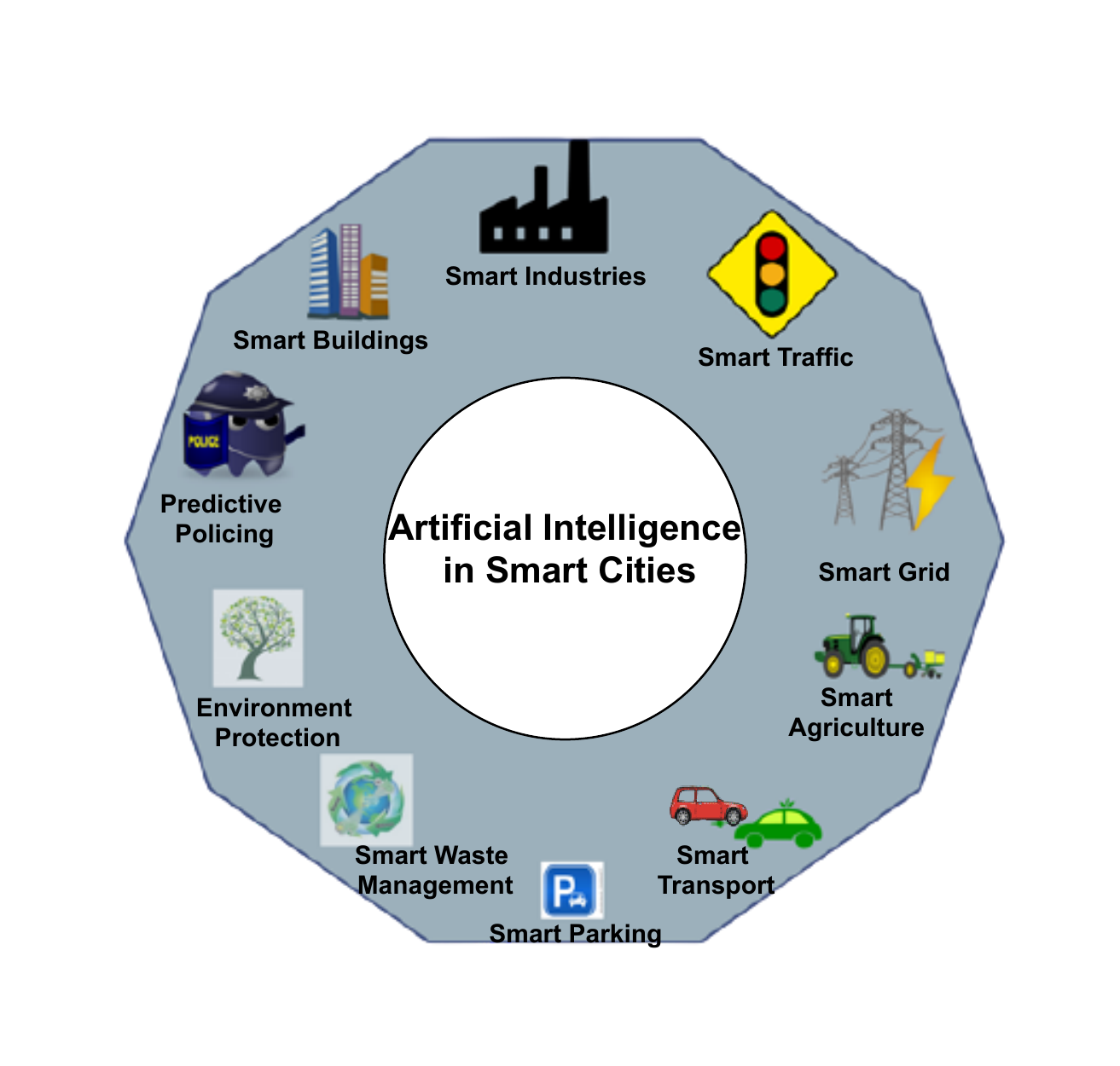}
\caption{Some interesting applications of AI in smart cities \cite{AI_applications_smart_city}.}
	\label{fig:applications}
\end{figure}

Some of the currently available key applications of AI in smart cities are illustrated in 
Figure \ref{fig:applications} and described next.
\begin{itemize}
 
 \item \textit{Healthcare}: The basic motivation of ML applications in healthcare lies in its ability to automatically analyze, identify hidden patterns, and extract meaningful clinical insights from large volumes of data, which is beyond the scope of human capabilities. The automatically extracted insights are generally efficient, and help medical staff in planning and treatment, ultimately leading to effective and low-cost treatment with increased patient satisfaction \cite{corbett2017real}. In recent years, AI has been heavily deployed in healthcare, and proved very effective, thanks to the recent advancement in deep learning. For instance, a solution proposed by Google \cite{Google} outperformed human doctors (i.e., by around 16\% accuracy) in the identification of breast cancer in mammograms. Similarly, AI solutions proposed in \cite{han2020keratinocytic,bhandary2020deep}, have been proved very effective in the diagnosis of skin and lung cancer, respectively. 
 
 \item \textit{Transportation and Autonomous Cars}: Transportation can benefit from AI in several ways. For instance, its predictive capabilities can help in traffic volume and congestion estimation for route optimization \cite{lee2020intelligent}. \textcolor{black}{AI algorithms can also be jointly used with multimedia processing techniques for road safety \cite{nguyen2019vehicle}, driver distraction \cite{li2021temporal}, and accident events detection \cite{bai2019traffic} and road passability analysis \cite{ahmad2019deep,ahmad2019automatic}.} However, AI can be considered as a backbone of autonomous cars where one of the key responsibilities of the AI module is continuous monitoring of the surrounding environment and the prediction of different events, which generally involves the detection and recognition of various objects such as pedestrians, vehicles, and roadside objects \cite{kuutti2020survey}. 
 
 \item \textit{Education}: AI brings several advantages in education by contributing to several tasks, such as automatic grading and evaluation, students' retention and dropout prediction, personalized learning, and intelligent tutoring systems \cite{ahmad2020artificial}. AI predictive capabilities could also help in predicting students' career paths by applying AI techniques on students' data covering different aspects, such as interests and performances in different subjects. 
 
 \item \textit{Crime Detection/prediction and Tracking}: It is another interesting smart city application where AI has shown its potential. AI is transforming the way law enforcement agencies operate to prevent, detect, and deal with crimes. In the modern world, law enforcement agencies are heavily relying on predictive analysis to track crimes and identify the most vulnerable areas of a city, where additional force and patrolling teams could be deployed. One example of such tools is PredPol, which relies on AI techniques to predict ``hot spot'' crime neighborhoods \cite{huet2015server}. 
 
 \item \textit{Clean and Sustainable Environment}: AI also helps in monitoring and maintaining a clean and sustainable environment. Thanks to the recent advancement in deep learning and satellite technologies, environment monitoring and enforcement are more efficient than ever \cite{Agriculture}. AI techniques have been widely deployed in analyzing remotely sensed data for environmental changes. \textcolor{black}{Moreover, AI techniques have also been demonstrated to be very effective in disaster detection \cite{said2019natural}, water management \cite{saboe2021real}, and waste classification \cite{ahmad2020intelligent}. }
 

 \item \textit{Smart Building}: Smart building represents an automatic structure/system to control the building's operations such as lighting, heating, ventilation, air conditioning, and security. AI has been widely exploited for various tasks in smart systems as elaborated upon in \cite{qolomany2019leveraging}.
 
 \item \textit{Tourism, Culture, Services, and Entertainment}: Tourism and entertainment industries are also benefited big time by AI and social media \cite{go2020machine}. For instance, AI-based recommendation systems are widely used by travelers in the decisions of their holidays' destinations considering different variables, such as transportation and accommodation facilities, cost, food, and historical points. In addition, AI-based applications could help travelers in fraud detection, cost optimization, and identification of entertainment venues and transportation facilities at the destination. Apart from the recommendation systems, which is one of the main applications of AI in the sector, AI enabled visual sentiment analysis tools could be used to search or extract scenes from long TV show videos based on sentiment analysis \cite{ahmad2020deriving}. 
 
\end{itemize}

Despite the outstanding performances and success, AI also brings challenges in the form of privacy and unintentional bias in public services. For instance, to analyze people's commute patterns, the administration needs to collect and process a lot of people’s data, including their movements risking people's personal information to be leaked. The intentional and unintentional bias in decisions of AI is even more dangerous, which might endanger citizens' lives in healthcare or law enforcement applications. For instance, an AI-based software used for future criminals predictions was found biased against blacks \cite{crawford2016artificial}. 
\textcolor{black} {Similarly, the smart system used to predict the health care needs of about 70 million patients in the US was assigning higher risks (scores) to black patient compared to white patients with the same medical conditions \cite{obermeyer2019dissecting}.}  It must be noted that the algorithms do not learn the bias on their own rather it comes from the data used to train the algorithms, which reflects the social and institutional biases of the society practiced over the years \cite{green2019smart}. Moreover, being a product of humans, AI algorithms reflect the beliefs, objectives, priorities, and design choices of humans (i.e., developers). For instance, to make accurate predictions, AI algorithms need the training data to be properly annotated and must contain sufficient representation for each class. An over-representation of a class may develop a tendency towards the class in predictions. A trade-off between false positives and false negatives is also very crucial for AI predictions. These limitations of AI hinder its way of overcoming social and political biases to achieve smart cities' true objectives. According to Green \cite{green2019smart}, AI algorithms in smart city applications are mostly influenced by the social and political choices of the society and authorities. Therefore, to ensure privacy and reduce bias of AI algorithms in human-centric applications, we need to discuss the need, goals, and potential impact of their decisions on society before deploying them. 

Moreover, there are several security threats to AI models in smart city applications, for instance, attackers can launch adversarial attacks on AI models to bias the decisions by disturbing their prediction capabilities. For example, an adversarial attacker might turn off an autonomous car on a high way, and ask for money to restart it. A more serious situation could be stopping a train on the platform just before the arrival of the next train \cite{thippeswamy2019guide}. Another challenge is the lack of interpretability---which results in humans being unable to understand the causes of an AI model’s decision. To deal with such risks involved in deploying AI in smart city applications, the concept of explainability and ethics in AI has been introduced. 

In the next sections, we provide a detailed overview and analysis of the potential security, robustness, interpretability, and ethical (data and algorithmic) challenges to AI in smart city applications.




\subsection{Scope of the Survey}

\begin{figure}[!tbh]
\centering
\includegraphics[width=.99\linewidth]{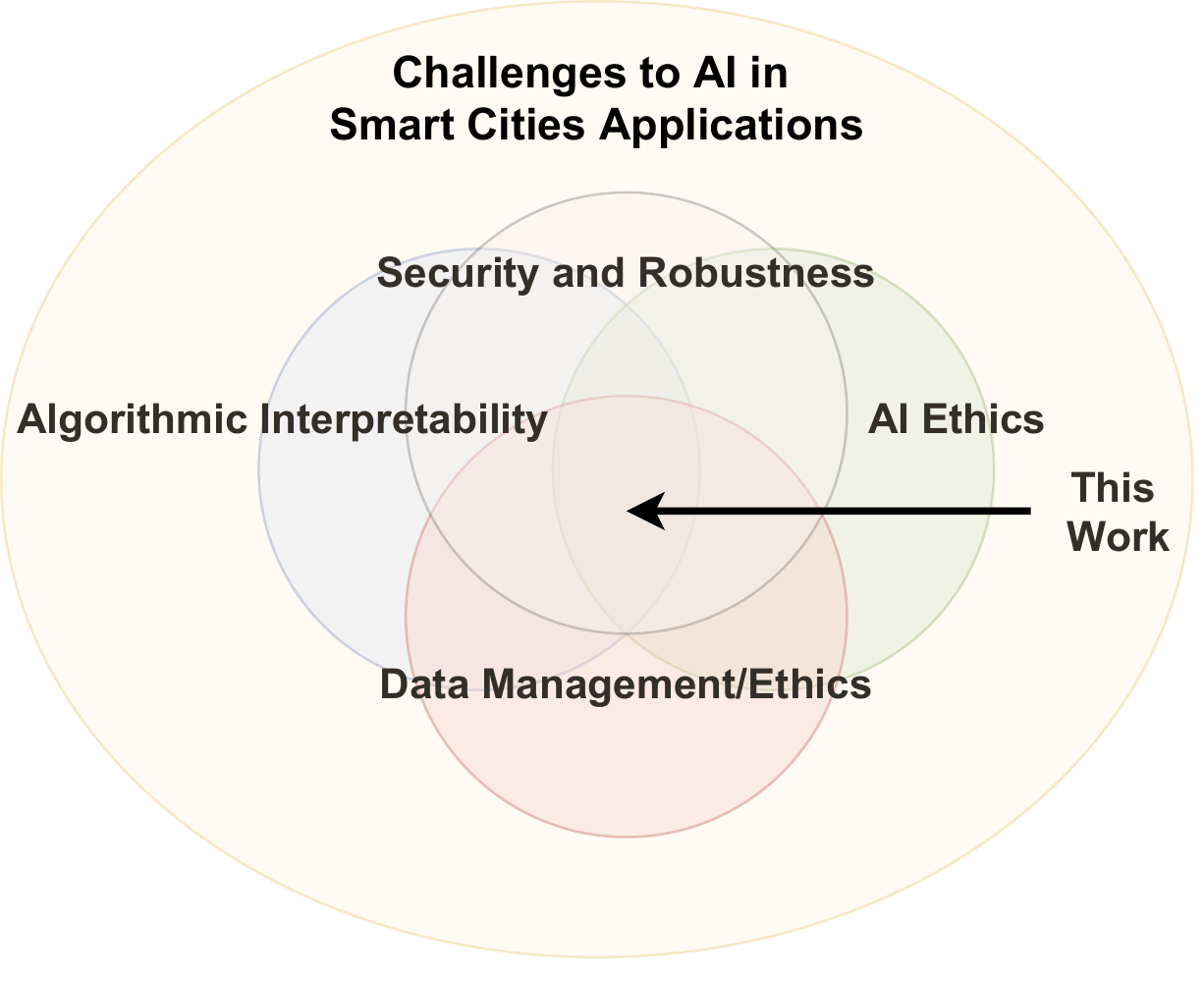}
\caption{Visual depiction of the scope of the paper.}
	\label{fig:scope}
\end{figure}

The paper revolves around the key challenges to a successful deployment of AI in smart city applications including security, robustness, interpretability, and ethical (data and algorithmic) challenges. Figure \ref{fig:scope} visually depicts the scope of the paper. The paper emphasizes these concepts/challenges by exploring how one of these challenges/problems may cause or help in solving others. We also analyze research trends in these domains. The paper also advises on the current limitations, pitfalls, and future directions of research in these domains, and how it can fill the current gaps in the literature, and lead to better solutions.

\begin{figure*}[]
\centering
\includegraphics[width=.89\linewidth]{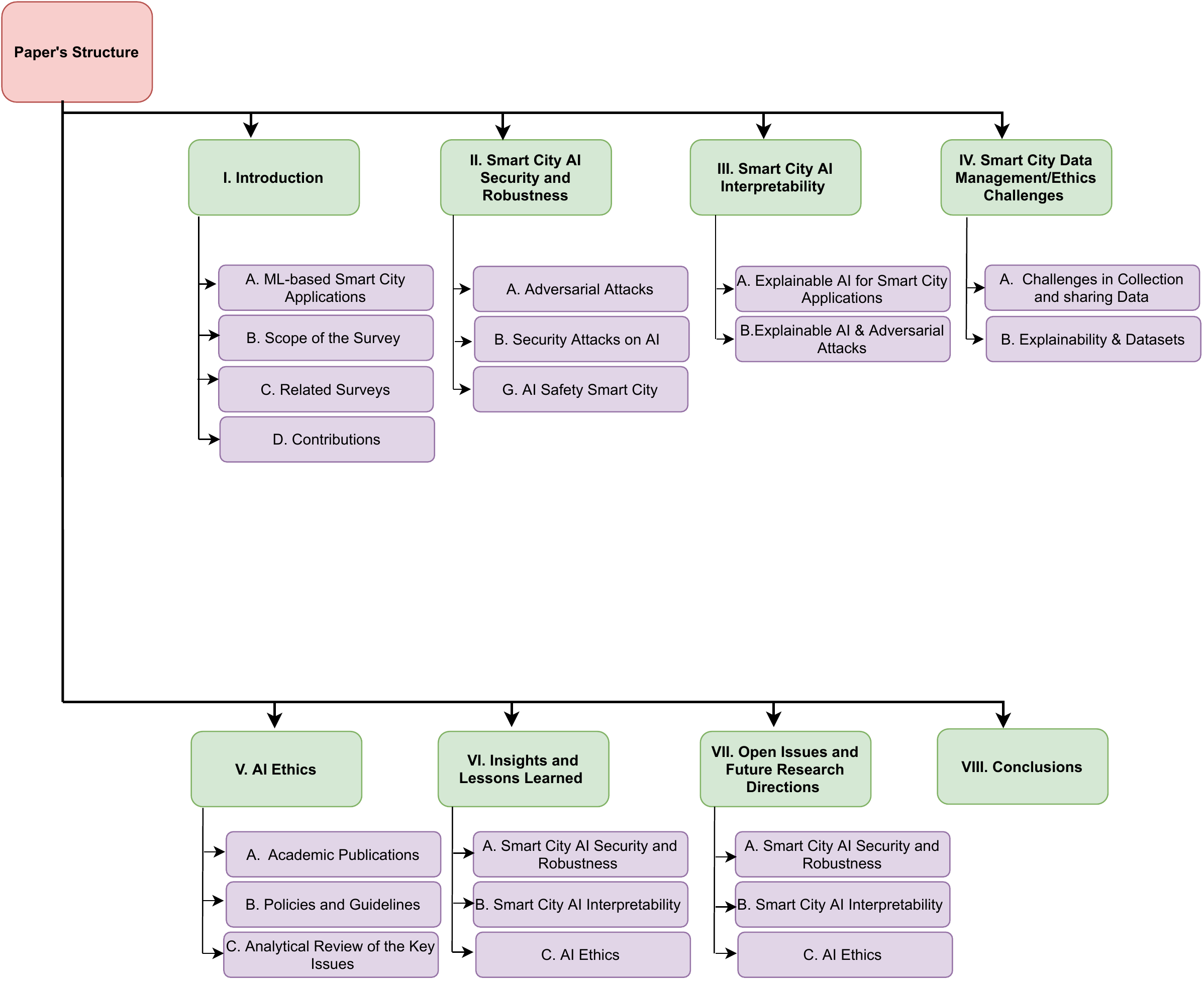}
\caption{Structure of the survey.}
	\label{fig:structure}
\end{figure*}
\subsection{Related Surveys}

Due to the keen interest in the research community in leveraging AI for smart city applications, it has been always a popular area of research \cite{mora2017first}. In literature, several interesting articles analyzing different aspects of AI applications in smart cities have been proposed \cite{ullah2020applications}. In addition, being among the key active research topics, a significant amount of the literature can be found on adversarial attacks, explainability, availability of datasets, and the ethical aspects of AI in human-centric applications. Adversarial and explainable AI are comparatively more explored in the literature. There are also some interesting surveys on these topics covering different aspects of individual topics. However, to the best of our knowledge, there is no survey jointly analyzing the challenges, and more importantly, emphasizing the connection between the four challenges. For instance, Zhang et al. \cite{zhang2020adversarial} provide a survey of adversarial attacks on deep learning models. Similarly, Serban et al. \cite{serban2020adversarial} provide a comprehensive survey on adversarial examples of object recognition. Zhou et al. \cite{zhou2019survey} on the other hand provide a survey of game-theoretic approaches for adversarial AI. There are also some recent surveys on explainable AI. For instance, in \cite{roscher2020explainable,tjoa2019survey,adadi2018peeking,arrieta2020explainable}, a survey of existing literature on explainable AI is presented. Some surveys focus on a particular type of technique for explainable AI. For instance, \cite{seeliger2019semantic} and \cite{puiutta2020explainable} survey web technologies and reinforcement learning-based approaches for explainability. Baum et al. \cite{baum2017survey} on the other hand, provides a survey of AI projects on ethics, risk, and policy. Similarly, Morley et al. \cite{morley2020ethics} provides an overview of the literature on AI ethics in healthcare. In contrast to other surveys, this paper emphasizes the connection between these four challenges, and analyze how a solution to one of the challenges may also help or cause the others. 
\subsection{Contributions}
In this paper, we provide a detailed survey of the literature on the security, safety, robustness, interpretability, and ethical (data and algorithmic) challenges to AI in smart city applications. The paper mainly focuses on the connection among these concepts, and analyzes how these concepts and challenges are dependant on each other.

The main contributions of the paper are summarized as follow: 

\begin{itemize}
 \item We provide a detailed analysis of how AI is helping in developing our cities, and the potential challenges, such as salient ethical, interpretation, safety, security, and fairness, hindering its way in different smart city applications.
 \item The paper analyzes the literature on major challenges including security, safety, robustness, interpretability, and ethical challenges in deploying AI in human-centric applications.
 \item The paper provides useful insights into the relationship among these challenges and describes how they may affect each other.
 \item We also identify the limitations, pitfalls, and open research challenges in these domains.
\end{itemize}

The rest of the paper is organized as follows. 
Section \ref{sec:security_and_Robustness} provides an overview of different security and robustness challenges to AI in smart city applications. Section \ref{sec:explainable_ML} details the importance of interpretability and explains how explainable AI can help in extracting more insightful information from AI decisions, and how it can be linked with adversarial attacks. Section \ref{sec:datasets} details the challenges associated with data collection and sharing. Section \ref{sec:ethics} focuses on the ethical aspects of deploying AI in human-centric smart city applications. Section \ref{sec:insights} summarizes the key insights and lessons learned from the literature. In Section \ref{sec:research_directions}, we highlights the open issues and future research directions. Finally, Section \ref{conclusion} provides some concluding remarks. \textcolor{black}{Figure \ref{fig:structure} visually depicts the structure of the paper.}

\section{Smart City AI Security and Robustness}
\label{sec:security_and_Robustness} 
Machine Learning has tremendous potential in smart cities that can improve the productivity and effectiveness of the different city systems. Despite the positive outcomes and the promise of AI in smart cities, security is one of the main concerns that still need further investigation and experiments. AI models can be vulnerable to different kinds of attacks; such as adversarial examples, model extraction attacks, backdooring attacks, Trojan attacks, membership inference, model inversion \cite{qayyum2020securing}. 

Attacks on AI models introduce new challenges to the existing software security systems and approaches that need to address a bit different nature of challenges \cite{ hussain2020machine}. AI has its unique security issues where a small modification on the objects (inputs or data consumed by AI algorithms) might change the decision of AI models and cause serious consequences. The following shortlist, on the security issues of AI applications in the last five years, clearly raises the urgent need to intensively study the safety and security aspects of AI while transforming cities to be smart.

\begin{itemize}
  
   \item 2016, the Auto-driver system in Tesla also confused the white side of a truck with the sky in 2016, leading to the deadly crash\footnote{Tesla website: \url{https://www.tesla.com/en_JO/blog/tragic-loss}}. 
   \item 2016, Microsoft chatbot was shut down and closed after a few hours of its release time. The model was attacked and forced to post offensive tweets for users\footnote{Microsoft: \url{https://blogs.microsoft.com/blog/2016/03/25/learning-tays-introduction/}}. \textcolor{black} {Chatbots aren’t just for businesses, but also for government services. The City of North Charleston in South Carolina, USA, has launched Citibot which is a communication tool for citizens and their governments. Citizens can ask for information or request repairs }\footnote{\url{https://www.northcharleston.org/connect/citibot/}}. \textcolor{black} {These smart systems are vulnerable to hacking as they consume data from citizens to analyze and manage their requests.} 
   \item 2016, Google AV was in autonomous mode where a failure in speed estimation caused a crash.\footnote{Media, technology news on BBC \url{https://www.bbc.com/news/technology-35692845}} 
   \item 2016, face recognition detection attack using eyeglasses frames \cite{sharif2016accessorize}. 
   \textcolor{black} {\item 2017, Apple face recognition has been fooled by a cheap 3D-printed mask\footnote{Forbes: \url{https://www.forbes.com/}}.} 
   
   \item 2018, Uber’s self-driving cars killed a pedestrian. The AV has not stopped at the right time.\footnote{The National Transportation Safety Board (NTSB), Media at: \url{https://www.theverge.com/2018/5/24/17388696/uber-self-driving-crash-ntsb-report}}
   \item 2018, robust physical perturbations could fool the DNN-based classifier of a self-driving car to misclassify speed limit signs \cite{eykholt2018robust}.
   \item 2018, \textcolor{black} {targeted audio adversarial examples have successfully attacked DeepSpeech, a deep learning-based speech recognition system.  AI works on sounds wave which can carry secret commands to the connected devices} \cite{carlini2018audio}. 
   \item 2019, the Tesla autopilot AI system has been attacked at Tencent’s Keen Security Lab by small changes on the lane markings, yet clear for humans. Tesla Model S swerve to the wrong lane making the lane recognition models in Tesla risky and unreliable under some conditions \cite{ackerman2019three}. 
   \item 2019, a neural network model diagnosis a benign mole as Malignant because of tiny noise added to the medical image\cite{finlayson2019adversarial}. 
   \item 2019, Deepfakes. Facebook creates a dataset for Deepfake detection.\footnote {Facebook AI: \url{https://ai.facebook.com/blog/deepfake-detection-challenge/ }}
   \textcolor{black} {\item 2019, the smart algorithm guiding care for tens of millions of people is biased against dark-skinned patients in New Jersey, USA. It is assigning dark-skinned patients lower scores than white patients with the same medical conditions.} \footnote{Media: \url{https://www.wired.com/}} 
  \textcolor{black} { \item 2020, A shopping guide robot in Fuzhou Zhongfang Marlboro Mall, China, walked to the escalator by itself. It fell off the escalator and knocked over passengers. The robot has been suspended from its duties\footnote{Media: \url{https://syncedreview.com , https://weibo.com/}}.
  }
  \textcolor{black} { \item 2020, Starsky Robotics has been shut down due to a safety issue in the self-driving software on highways. The team of Starsky reported that supervised ML is not solely enough to build a safe robot-trucks industry\footnote{Medium:  \url{https://medium.com/}}.
  }
  \item \textcolor{black}{2021, Tesla cars crash due to autopilot feature\footnote{https://www.jumpstartmag.com/ai-gone-wrong-5-biggest-ai-failures-of-all-time/}.}
  \item \textcolor{black}{2021, FBI issued a warning about a rise in AI-based synthetic materials including deepfake content\footnote{https://securityintelligence.com/articles/how-protect-against-deepfake-attacks-extortion/}.}
    \item \textcolor{black}{2021, AI chatbot has been suspended, and the firm is now being sued in South Korea, for making offensive comments and leaking users information.\footnote{https://www.straitstimes.com/}.}
 \end{itemize}
 
This list indicates that there are several issues to be handled beyond building AI models of good performance. In the following subsections, we discuss the strategies of attacks on machine learning models in smart city applications which shed the light on the necessity for safe and robust AI solutions at both technical and policy levels.

\subsection{Adversarial Attacks }
\label{sec:adversarial_attacks}
This challenge has been recognized and discussed either for crafting fake data that could belong to different domains; text \cite{sato2018interpretable}, images \cite{papernot2016limitations}, audio \cite{carlini2018audio}, network signals \cite{corona2013adversarial} known as adversarial examples or evaluating and developing solutions against this security threat  \cite{carlini2019evaluating}. Formally, given a benign input data $X$ which is classified as class 1 by model $M$, find a function $F$ to generate $X^{\prime}$ (poisoning function $F$; $F(X)$ = $X^{\prime}$) so that $X^{\prime}$ is classified as class 2 by the same model $M$ where the difference between $X$ and $X^{\prime}$ is not being discovered by humans. 

\begin{figure}[!h]
\centering
\includegraphics[width=.99\linewidth]{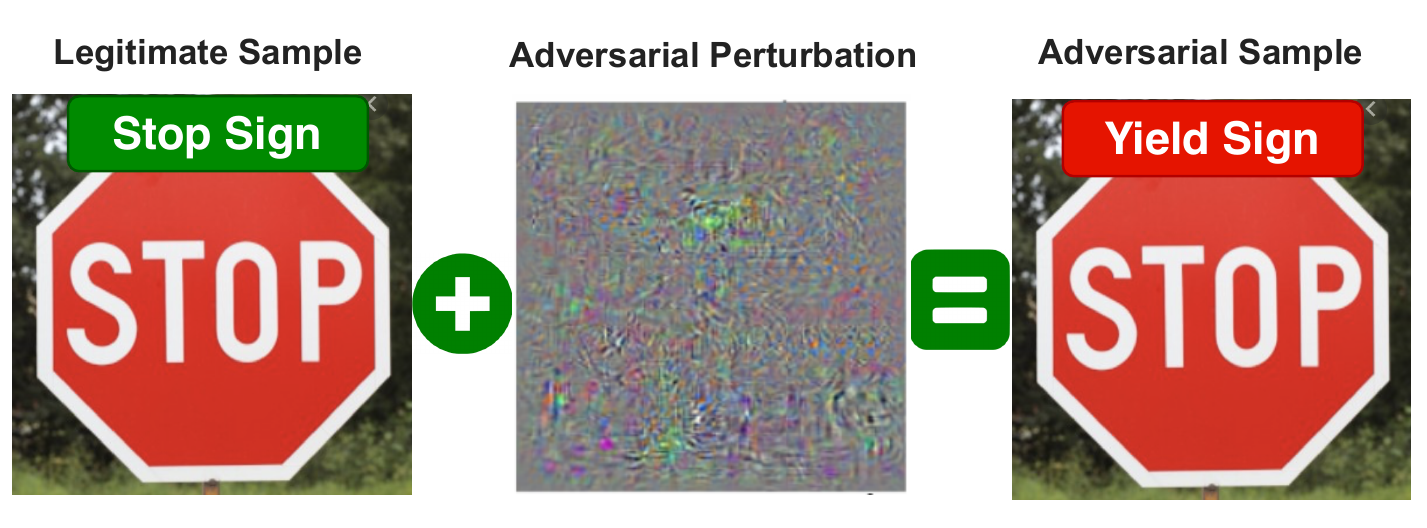}
\caption{An illustration of a common adversarial attack on image classification AI model. The shown adversarial perturbation (amplified for illustration) is added into a sample to force the model to make a wrong prediction with high confidence \cite{massoli2020cross}.}
	\label{fig:adversarial_example}
\end{figure}

This former definition is referred to as un-targeted adversarial attacks. The targeted attacks have a target class Y where function $F$ is trying to find another version of any benign input where model $M$ becomes biased towards $Y$ in prediction. Another classification of adversarial attacks is based on the amount of knowledge the attackers have about the target model (victim). The threat model can be a white-box, gray-box, or black-box. In white-box attacks, adversaries have full knowledge about the targeted model architecture. This eases the process of crafting poisoned data and thus fools the system. While in gray-box threat models attackers could have some information about the overall structure of the model, in black-box threat models all that they have is just access to use the model \cite{ ren2020adversarial}. A detailed taxonomy of adversarial attacks can be found in \cite{qayyum2020secure}. Figure \ref{fig:adversarial_example} illustrates a common adversarial attack on an image classification classifier, where an AI model has been deceived by adding a tiny perturbation, amplified in the figure for visual depiction, to a legitimate sample to disturb the prediction capabilities of the model.

Since the main focus of the paper is on smart city applications, thus without going into further details, in the next subsection, we provide an overview of the literature on adversarial attacks on AI models in smart city applications.

\subsubsection{\textbf{Adversarial AI and Smart City Applications}}
Adversarial attacks are considered severe security threats in learner-based models due to its possible consequences. In smart cities, complex networks, and collaborations of data-driven applications and devices, the impact of misleading a model, e.g., a classifier, could result in harsh situations and a costly mess. This could happen no matter the attack has intentionally misled the model such as crafted inputs by attackers, or unintentionally ``accidentally'' such as a defect in traffic light signals, or varying weather conditions that could impact signs illumination consumed by autonomous vehicles \cite{stallkamp2012man}. 
In \cite{papernot2017practical}, a perturbation on a regular image of a stop sign forces a deep neural network classifier to see it as a yield sign. This information can lead the vehicle to behave unsafely and might cause severe accidents. This case could be worse if other neighboring vehicles consume some data sent by the attacked vehicle. A DNN-based solution was developed in \cite{ khanapuri2019learning} to detect and then isolate the attacked vehicle from the cooperative vehicles. AI models in Autonomous vehicles depend not only on the exchanged sensors data but also on consuming street signs to control the driving and traffic. The security of these models is crucial since a slight change in sign image could be enough to fool the model, for example, one pixel is often enough to attack a classifier in \cite{su2019one}.

Similarly, human lives and billions of dollars could be victims of AI models that are misclassifying the diseases and medical reports. In \cite{finlayson2019adversarial} using a slight noise on disease images or even replacing some words in disease description by their synonyms, the AI models changed the decisions to the opposites of the true ones.
Despite that medical images are taken in pre-defined settings, where some manipulations applied to other domains images are not valid such as rotations, some manipulation methods can be easily detected by specialists eyes \cite{taghanaki2018vulnerability}, there is still a chance to be manipulated by other methods \cite{ finlayson2018adversarial}. In \cite{ becker2019injecting}, GAN was able to modify breast medical images through adding/removing features and change the AI decision while radiologists never discriminate the difference between the original and manipulated images at low-resolution rates. Brain medical images have been manipulated by three different methods; noise generated, fast gradient sign, and virtual adversarial training to generate adversarial examples to mislead the brain tumor classifier  \cite{kotia2019risk}. Fortunately, with the help of DNN, a detector has been developed and showed surprising high accuracy in detecting manipulated medical images \cite{ma2020understanding}. Another detector is a result of ensemble CNN networks \cite{paul2020mitigating}, or by training dataset augmentation by the adversarial examples of modified CT scan images \cite{liu2020no}
The literature shows more evaluation of medical image attacks than text attacks. This is probably because the attacks arise in the computer vision field. However, texts in natural language are also liable to attacks \cite{ zhang2020adversarial}. This means prescriptions, medical records classification for insurance decisions, patient history, and allergic information, medical claims codes that determine reimbursement, are all vulnerable to attacks. The sensitive nature of these applications and the resulting harms (economic and well as social) raises the concern for safety, security, and dependability of AI systems. In the future, extra computational interventions (e.g. adversarial data detectors) may form an integral part of AI-based medical solutions.

Other components of smart cities are not far from serious attacks. In the smart energy sector, attacks come in different forms; denial-of-service where systems or part of them become inaccessible and can also be optimized for more sufficient energy needs \cite{ zhang2016attack}, randomly manipulate the sensor readings, or with some information the attacker has about the system and sensors, false data are injected to the system \cite{manandhar2014detection, marulli2019adversarial, chen2019exploiting}. Several detection solutions have been proposed and evaluated to mitigate the attacks in a grid such as false data injection detection \cite{ beg2017detection}, securing the gird physical layers against attacks \cite{ islam2019physical}.

Adversarial attacks also have a serious impact on food safety and production control \cite{ fawaz2019adversarial}. Several AI solutions feed on images, videos, text in smart agriculture, and smart waste. These two smart sectors may be more vulnerable to unintentional attacks, one reason is because of natural conditions where the sensors and cameras work. Table \ref{summary_adversarial_attacks} provides a summary of some of the works on adversarial AI in smart city applications.

\begin{table*}[tb!]
\centering
\caption{Summary of some key works in adversarial AI in terms of smart city application, type of attack (white/black-box), dataset and key features of the method.}
\label{summary_adversarial_attacks}
\scalebox{0.8}{
\begin{tabular}{|p{.5cm}|p{1.8cm}|p{2.5cm}|p{2.2cm}|p{2.5cm}|p{10.6cm}|}
\hline
\multicolumn{1}{|c|}{Ref.} & \multicolumn{1}{c|}{Application} & \multicolumn{1}{c|}{Type of attack} & \multicolumn{1}{c|}{\textcolor{black}{AI Model}} & \multicolumn{1}{c|}{Dataset} & \multicolumn{1}{c|}{Description of the method} \\ \hline
 \cite{sitawarin2018darts}& Transportation & White-box and Black-box &\textcolor{black}{ CNNs} & GTSRB \cite{stallkamp2012man} and GTSDB \cite{houben2013detection} & It proposes adversarial attacks on the traffic sign recognition systems/models of autonomous cars. It mainly proposes two types of attacks namely (i) Out-of-Distribution attacks, and (ii) Lenticular Printing attacks. The former modifies the innocuous signs in a way that the model predicts it as potentially dangerous traffic signs while the latter relies on an optical phenomenon to deceive the traffic sign recognition system. \\ \hline
 \cite{cao2019adversarial} & Transportation & White-box & \textcolor{black}{DNNs} & KITTI \cite{geiger2013vision}& It introduces adversarial attacks on LiDAR-based perception in autonomous vehicles where LiDAR spoofing attacks are used for generating fake obstacles in front of the target autonomous vehicle to disturb its decision-making abilities. \\ \hline
 \cite{li2020adaptive}& Transportation & Black-box & \textcolor{black}{DNNs} & GTSRB \cite{stallkamp2012man} & It targets Deep Neural Networks (DNNs) based traffic sign recognition model with black-box attacks by employing an efficient sampling strategy for Adaptive Square Attack (ASA) able of generating perturbations for traffic sign images with fewer query times. \\ \hline
\cite{finlayson2018adversarial} & Healthcare & White and black-box attacks & \textcolor{black}{CNNs} & Chest X-ray \cite{wang2017chestx} & The work demonstrates how adversarial attacks can be launched against deep learning based systems for healthcare. Moreover, it also analyzes how healthcare is susceptible to adversarial attacks both in terms of monetary incentives and technical vulnerabilities. \\ \hline
\cite{champneys2020vulnerability} & Healthcare & White-box &\textcolor{black}{MLP} & Los Alamos National Laboratory (LANL) dataset \cite{farrar1999structural} & It explores and demonstrates the vulnerabilities of data-driven approaches to structural health monitoring by generating/mapping the records of Los Alamos National Laboratory into adversaries using a white-box attack. Moreover, the work also proposes an adversarial threat model specific to structural health monitoring. \\ \hline
 \cite{newaz2020adversarial}& Healthcare & White and black-box & \textcolor{black}{DT, RF, and ANNs}& Self-collected & It proposes a new type of adversarial attacks for targeting AI models in healthcare where the attacker/adversary has partial information about the data distribution and AI model. The attacks are intended to change medical device readings to alter patient status/diagnosis results. \\ \hline
\cite{li2019desvig} & Industry & White and black-box &\textcolor{black}{ DNNs}& Self-collected & Proposes a decentralized framework namely DeSVig to identify and to guard against adversarial attacks on an industrial AI system. The biggest advantage of the framework is its ability to reduce failure of identifying and being deceived by adversaries. \\ \hline
\cite{zhang2020detecting} & Smart Grids & White-box & \textcolor{black}{GANs}& Self-collected & Explores the vulnerabilities in smart grids. To this aim, a data-driven learning-based algorithm has been proposed to detect un-observable false data injection attacks in distribution systems. Moreover, the method needs less training samples and makes use of unlabelled data in a semi-supervised way.\\ \hline
 \cite{chen2019exploiting} & Smart Grids & Black-box& \textcolor{black}{RNN and LSTM}& Self-collected & The work analyzes and explores the vulnerabilities of smart grids against adversarial attacks. The authors mainly focus on the key functions of smart grids, such as load forecasting algorithms, and analyze the potential impact of the adversaries on load shedding and increased dispatch costs using data injection attacks. \\ \hline
 \cite{zhou2019evaluating}& Smart Grids & Black-box & \textcolor{black}{RNN and LSTM}& Self-collected & The paper analyzes the vulnerabilities and resilience of AI models in power distribution networks against adversarial attacks on smart meters via a domain-specific deep learning architecture. Smart meters are attacked under the assumption that the attacker has full knowledge of both the model and the detector.\\ \hline
\cite{wang2020transferable} & Person Re-identification (Surveillance) & Black-box attacks &\textcolor{black}{CNNs} & Market1501 \cite{zheng2015scalable}, CUHK03 \cite{li2014deepreid}, and DukeMTMC \cite{ristani2016performance} & The work aims to explores and analyzes how the person re-identification in CCTV cameras frameworks can suffer from adversarial attacks. To this aim, the authors launch back-box attacks using a novel multi-stage network architecture stacking the features extracted at different levels for the adversarial perturbations. \\ \hline
\cite{fawaz2019adversarial}& Food safety & White-box & \textcolor{black}{CNNs} & UCR \cite{dau2019ucr}& The paper analyzes the vulnerabilities of deep learning algorithms in time-series data by adding noise to the input samples to decrease a deep learning model's confidence in food safety and quality assurance applications. \\ \hline
\end{tabular}}
\end{table*}
\subsection{Security Attacks on AI}
\textcolor{black}{In this section, we introduce the readers to some other common strategies to launch attacks on AI particularly in cloud and edge deployments, such as data poisoning, evasion attacks, exploratory attacks,  model extraction, backdooring, trojan, model-reuse, cyber kill chain–based attacks, membership inference, and model inversion attacks,  which are very common in smart city applications.}

 \subsubsection{\textbf{Data Poisoning}}
\begin{figure}[!h]
\centering
\includegraphics[width=.80\linewidth]{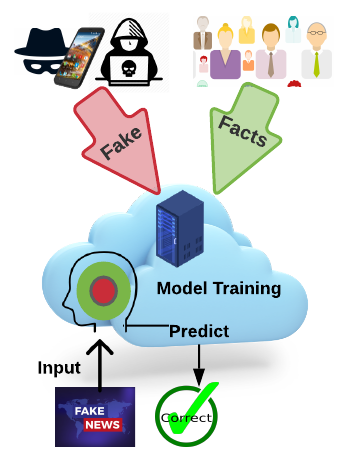}
\caption{The demonstration of the poisoning attack on machine learning, where the adversary can deliberately inject bad data (poisoned data) into the training pool which is then used for training models. Models are built on something that should not learn allowing subsequent mispredictions. }
	\label{fig:DPS-MM}
\end{figure}

In this attack, as illustrated in Figure \ref{fig:DPS-MM}, attackers intentionally share manipulated data, e.g. incorrect labels, so the model would consume in any re-training process with a target to degrade the AI models’ performance. In this case, attackers somehow have control over the training data or can contribute to the training data \cite{dunn2020robustness}. In smart cities, crowdsensing is an integral data source for smart services which is involved in several areas such as transportation, pollution monitoring, and energy management \cite{alvear2018crowdsensing}. However, it is highly susceptible to data poisoning attacks \cite{li2019deep,huang2019robust}, and in some settings, gain greater degrees of reliability so that they are hard to be identified \cite{miao2018towards,miao2018attack}. In a very sensitive field of study, an experiment on around 17,000 records of healthy, unhealthy (disease-infected) people, a poisoning attack on the training data was able to drop the classifier accuracy of about 28\% of its original accuracy by poising 30\% of the data  \cite{newaz2020adversarial}. This could have severe consequences, for example, on dosage or treatment management.

\subsubsection{\textbf{Evasion Attacks}}
\begin{figure}[!h]
\centering
\includegraphics[width=.9\linewidth]{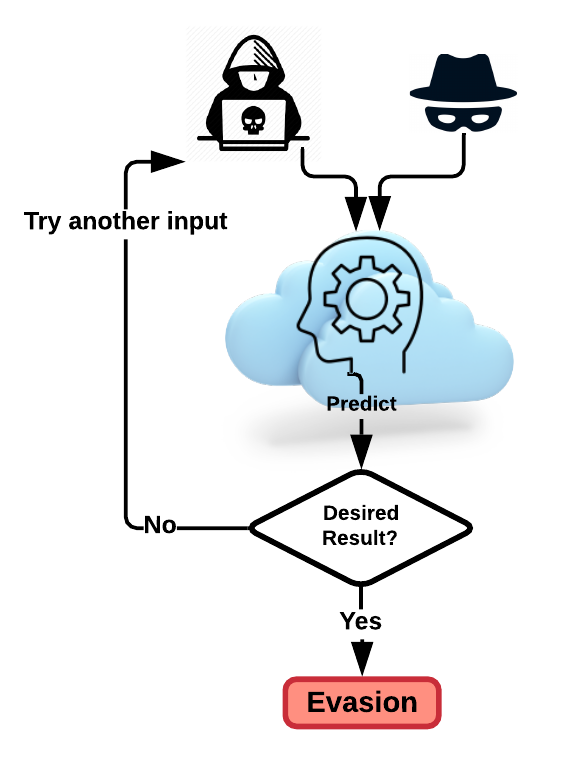}
\caption{The demonstration of the evasion attack on machine learning, where the adversary queries the model by carefully crafted examples, yet seems normal for a human, to have them misclassified. The attackers try to perturb the data in iteration mode where a little noise is added in each iteration until the input changes its original label according to the model.}
	\label{fig:EVAS-MM}
\end{figure}

Compared to data poisoning, evasion attacks can take place after model training as shown in Figure \ref{fig:EVAS-MM}. The attackers may have no idea about the required data manipulation to attack the model. A practical evaluation in \cite{biggio2013evasion} shows that commonly used classification algorithms, such as SVM and NN, can be easily evaded even with limited knowledge about the system, and an artificial dataset. To highlight the risk of using deep learning in the context of malware detection, \cite{demetrio2019explaining} proposed a novel attack that changes a few bytes in the file header without injecting any other data and forces MalConv, a convolutional neural network for malware detection, to misclassify the benign and fabricated inputs. In this attack strategy, attackers would keep querying the AI models in a trial and error fashion so they can learn how to design their inputs to pass the model. This would create an overhead on the systems and a solution to identify suspicious queries that may save the availability of the systems and the power consumption, especially if the target models run on devices of limited energy supply. 

\subsubsection{\textbf{Trojan Attacks}}
\begin{figure}[!h]
\centering
\includegraphics[width=.79\linewidth]{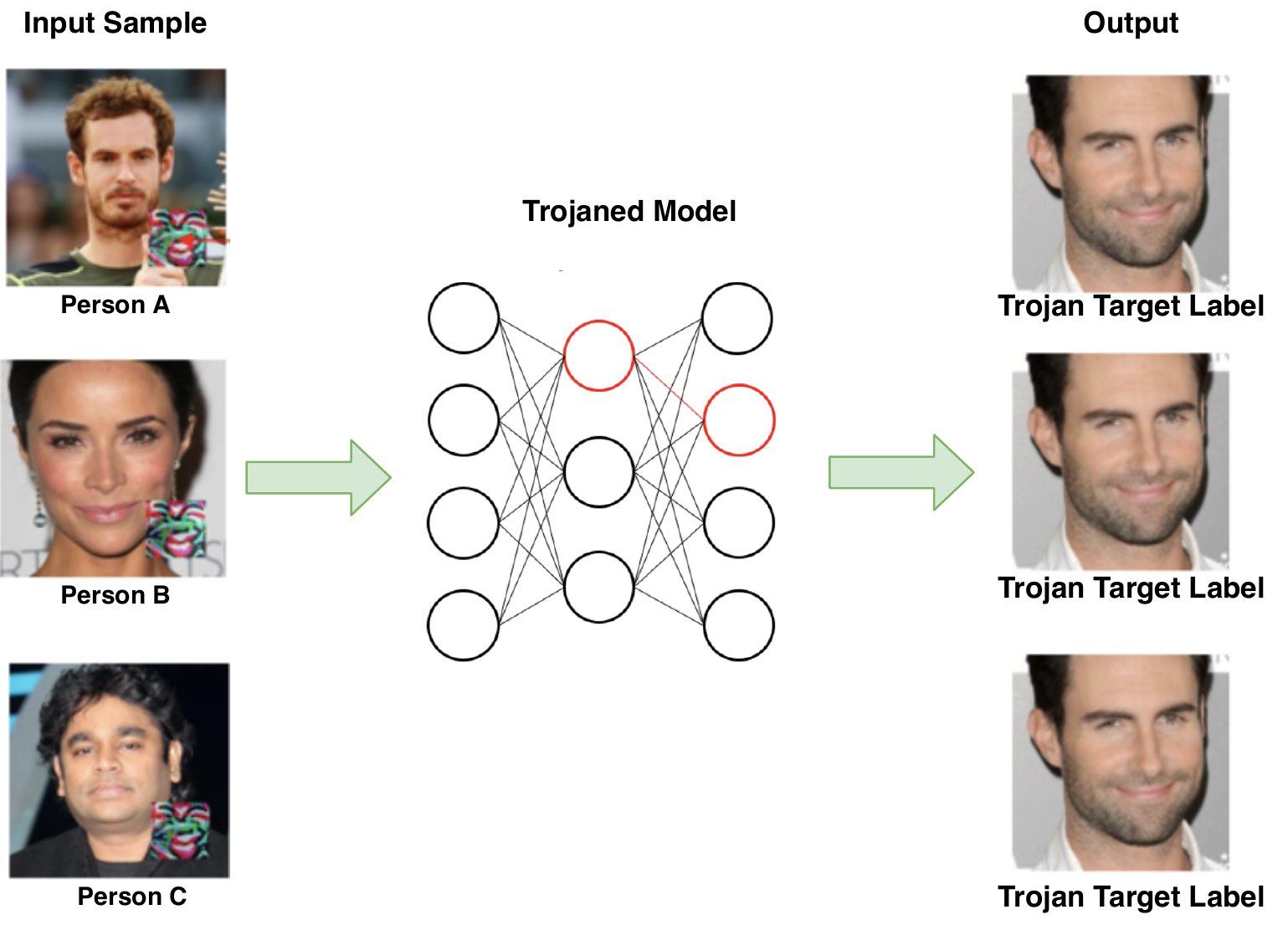}
\caption{An illustration of a trojan attack on face classification AI model. The trojan trigger added into a sample activates the trojan attack and predicts/generates the trojan target label \cite{massoli2020cross}.}
	\label{fig:trojan_example}
\end{figure}

Trojan attacks on AI algorithms are also very common in cloud and edge deployments of AI \cite{liu2020survey,gao2019strip}. In a trojan attack, the attackers modify the weights of a model in a way that its structure remains the same. Moreover, a trojan attacked AI model works fine on normal samples, however, it predicts the trojan target label for an input sample when a trojan trigger, which is an infected sample to activate the attack, is activated. Figure \ref{fig:trojan_example} illustrates how trojan attacks behaves when a torjan attack is triggered on a face recognition system. In this case, the victim classifier always predicts the trojan target label when test samples with trojan trigger are used. 

\subsubsection{\textbf{Model Stealing (Model Extraction)}}
\begin{figure}[!h]
\centering
\includegraphics[width=.9\linewidth]{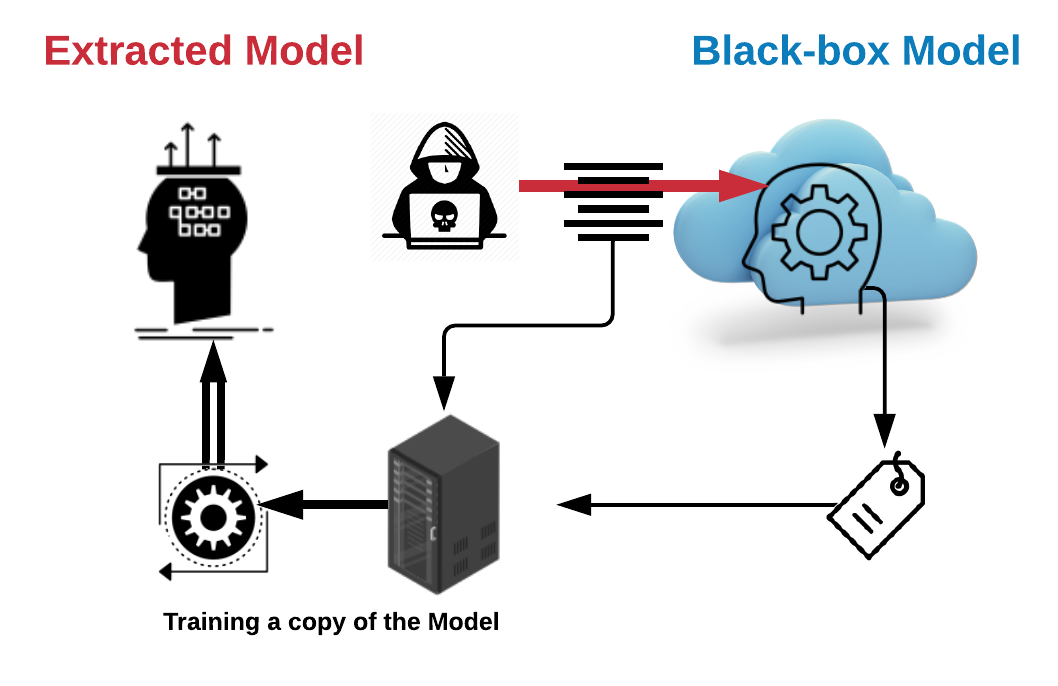}
\caption{The demonstration of the model extraction attack, or stealing, in machine learning. In such an attack, the adversary queries the classifier by different inputs and collects the labels. The combination of the returned labels and the input data is used to build a training dataset to train another model.}
	\label{fig:MES-MM}
\end{figure}

This strategy is also called model extraction as illustrated in Figure \ref{fig:MES-MM}. As its name implies, the ultimate objective of the adversary is to clone or reconstruct the target model, re-engineering a black-box model, or to compromise the nature and the properties of the training data \cite{juuti2019prada}. This strategy of attacking the AI models dated back to 2005 when the authors of \cite{lowd2005adversarial} were able to develop an effective algorithm for reverse engineering a spam filter model. Compared to the above two attack strategies in AI applications, model extraction needs neither any knowledge about the training data nor the model properties and architecture. All that the adversaries have is access to the model and they get its answers to the submitted queries \cite{orekondy2019knockoff,papernot2017practical}. The MLaaS could be the main target of this attack since a few dollars may help in creating a free copy of the paid model over cloud \cite{ krishna2020thieves}. Creating a private copy of the victim model not only a copyright issue but also expose the victim model to other attacks of different strategies since the attackers have new information on crafting adversarial examples \cite{ papernot2017practical,orekondy2019prediction}.

\subsubsection{\textbf{Membership Inference Attacks}}

\textcolor{black}{In such attacks, the attackers do not necessarily need knowledge about the parameters of an AI model rather a knowledge of the type and architecture of the model and/or the service used for developing the model is used to launch an attack. Such attacks are very common due to the growing interest in using AI as a service allowing the attackers to develop and launch membership inference attacks using the same services. For instance, Shokri et al. \cite{shokri2017membership} proposed a membership attack technique capable of launching attacks on AI models developed using Amazon and Google services.} 

\textcolor{black}{The severity and risks associated with membership inference attack largely depend on the applications and the type of data used for training an AI model. In certain applications involving complex image and speech classification tasks, the efforts involved in generating training data reduce the severity of the attacks. On the other hand, in some human-centric applications, such as education, finance, and healthcare applications with tabular data, which can be easily generated, membership inference attacks may have server implications.}

Table \ref{summary_security_attacks} provides a summary of some of the works on security attacks on AI in cloud and edge deployment for smart city applications.
\begin{table*}[tb!]
\centering
\caption{Summary of some key works on security attacks on AI in cloud and edge deployment in terms of application, type of attack, dataset and key features of the method.}
\label{summary_security_attacks}
\scalebox{0.8}{
\begin{tabular}{|p{.5cm}|p{2.5cm}|p{2cm}|p{1.5 cm}|p{3cm}|p{10.6cm}|}
\hline
\multicolumn{1}{|c|}{Ref.} & \multicolumn{1}{c|}{Application} & \multicolumn{1}{c|}{Type of attack}& \multicolumn{1}{c|}{\textcolor{black}{AI Model}} &
\multicolumn{1}{c|}{Dataset} & \multicolumn{1}{c|}{Description of the method} \\ \hline
 \cite{liu2018sin}& Fashion and supply chain& Trojan &\textcolor{black}{DNNs} & Fashion-MNIST \cite{xiao2017fashion} & It targets DNNs using stealth infection on the models in a cloud-based neural computing frameworks. It is to be noted the attack aims to harm the end users without any impact on the service provider. \\ \hline
 \cite{davaslioglu2019trojan} & Communication & Trojan& \textcolor{black}{CNNs}& RML2016.10A \cite{o2016radio} & It introduces and launches Trojan attack against a AI model for wireless signal classification. Moreover, the authors evaluate several types of mechanisms for the detection of such attacks on wireless signal classification. \\ \hline
 \cite{gao2019strip}& Autonomous Cars & Trojan & \textcolor{black}{DNNs}& GTSRB \cite{stallkamp2012man} & It proposes and develops a framework namely STRong Intentional Perturbation (STRIP) to guard against run-time Trojan attacks on image classification frameworks where certain input samples are intentionally perturbed by superimposing various image patterns to analyze the behavior of the model for the detection of malicious samples. \\ \hline
 \cite{shi2018spectrum}& Communication & Data Poisoning (exploratory attack) &\textcolor{black}{DNNs} & Spectrum Sensing \cite{davaslioglu2018generative} & It proposes a data poisoning attack on an AI model for the cognitive transmitter by changing the channel occupancy information, for instance from busy to idle or vice-versa, to disturb the decision capabilities of the model on certain samples. \\ \hline
 \cite{steinhardt2017certified}& Sentiment Analysis & Data Poisoning &\textcolor{black}{DNNs} & IMDB \cite{yenter2017deep} &  It guards against data poising attacks on AI models by constructing approximate upper bounds on the loss under two assumptions: (i) the `` dataset is large enough for a statistical concentration between train and test error to hold,'' and (ii) the outliers in the cleaned dataset does not have any impact on the model. The technique is meant to the defender aiming outlier detection. \\ \hline
\cite{kesarwani2018model} & IRS tax pattern and Email Importance & Model Extraction & \textcolor{black}{DT} & IRS Tax Pattern, GSS Survey, Email Importance, and Steak Survey \cite{yenter2017deep} & It provides a mechanism for guarding against extraction attacks where a framework is firstly attacked with extraction attacks by measuring the learning rate of the model. A cloud-based extraction monitoring mechanism is then developed to quantify the extraction status of models by analyzing the query and the corresponding response streams.  \\ \hline
 \cite{correia2018copycat} & Facial Expression Recognition & Model Extraction &\textcolor{black}{CNNs} & Multiple dataset including AR Face \cite{martinez1998ar}, BU3DFE \cite{yin20063d}, and JAFFE \cite{lyons1998coding} & It aims to analyze whether a black box CNNs model can be steal or not? To this aim, a CNN model is queered with unlabeled samples to extract the model's information by analyzing its response to the unlabeled samples, which are used to create a fake dataset then. \\ \hline
 \cite{hitaj2018have}& Digits Classification & Evasion Attacks  &\textcolor{black}{CNNs} & MNIST \cite{mizukami2010cuda} & It analyzes the robustness and reliability of one of the commonly used types of evasion attacks defense methods namely watermarking schemes for CNNs where the authors claim that attackers can evade the verification of original ownership under such schemes.  \\ \hline
\end{tabular}}
\end{table*}
\subsection{AI Safety in Smart City}
The concept of safety in AI is not much different from its definition in other engineering sectors. It mainly covers the minimization of risk and uncertainty of damage \cite{mohseni2019practical}. 

An important issue is that the safety evaluation of the AI-based systems could need further effort beyond the testing dataset since the real environment could have a larger probability of uncertainty and risk. The models trained and tested on large datasets could be more robust in production environments  \cite{varshney2017safety}. This simply means the availability of useful and representative datasets not only a concern to get benefit from AI algorithms, but also to build more safe and robust solutions. In subsequent sections, we discuss the challenge of dataset availability.

Unsafe machine learning-based solutions could impact the lives of creatures directly; such as those systems in AV that killed people, or indirectly by raising racist issues, for example. The serious issue of Tesla auto-driver is the human driver was killed because of a mistake after millions of miles in testing the auto-driver system. The Google photo app also returned racist results after training on thousands of images. This simply means, even with the availability of massive datasets, AI-based systems still need serious and solid research works to mitigate the effect of mistakes and develop counter-strategies against illegal usage; adversarial attacks for example.

\section{Smart City AI Interpretability}
\label{sec:explainable_ML}
In a typical AI framework, a set of features is feed to an AI algorithm, which learns from the data by identifying a hidden pattern, and in return produces some predictions. In such frameworks, which are also termed as black-boxes, the predictions come without any justification/explanation, and the users have no idea of the reasons behind the outcome. On the other hand, in an explainable AI framework, besides prediction/decisions, an AI model also details the causes of the prediction/decision. To this aim, additional functions/interface is used to interpret the causes behind an underlying decision \cite{arrieta2020explainable}. \textcolor{black}{In the literature, interpretability and explainability are generally used interchangeably. However, the terminologies are relevant but slightly different. Interpretability shows the extent to which a cause and effect can be observed within a system while explainability represents the extent of explanation/description of AI algorithms mechanism to a human.  }

There are several factors motivating the need for explanation/justification of the potential causes of an AI model in general and in smart city applications in particular, where justification and explanation of an AI model's outcome are very critical for developing the users' trust in AI models used to take some critical decisions about their lives, such as whether we get a job or not (AI-based recruitment), whether an individual is guilty/involved in a crime or not (i.e., predictive policing) etc., \cite{o2016weapons}. According to Guidotti et al. \cite{guidotti2018survey}, these justification of the causes of an AI model's predictions could be obtained in two ways either by developing techniques/methods to describe the potential reasons behind the model's decision, which they termed as ``Black-box Explanation'' or directly designing and developing transparent AI algorithms.

Some key advantages of explainable AI are:

\begin{itemize}
 \item Explainability of AI models helps in building users' trust in the technology, which will ultimately speed up its adoption by industry.
 \item Explainability is a must characteristic for AI models in some sensitive smart city applications, such as healthcare and banking.
 \item Explainable AI models are more impactful compared to traditional AI models in decision making.
 \item Helps in detecting algorithms' biases. 
\end{itemize}

\begin{figure*}[!h]
\centering
\includegraphics[width=.99\linewidth]{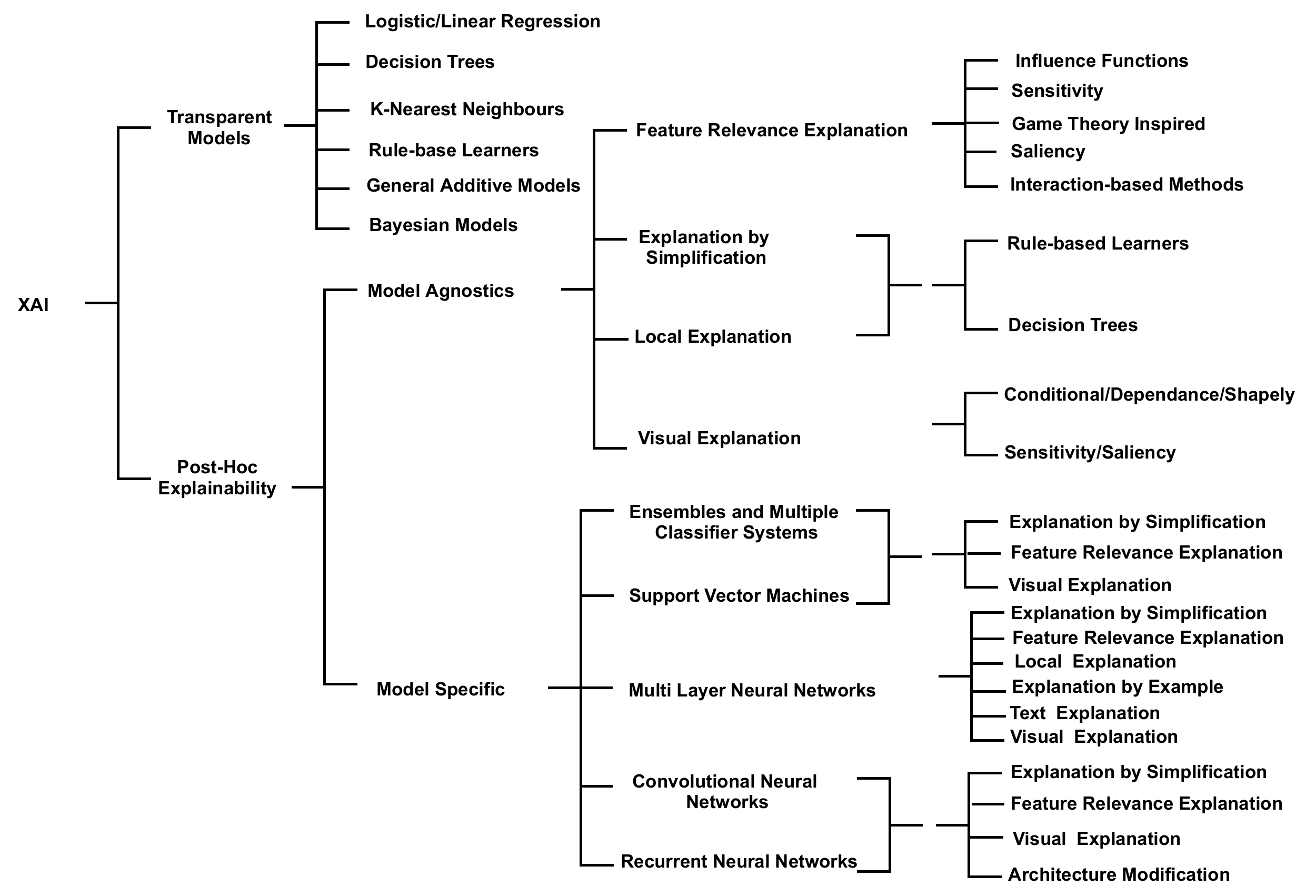}
\caption{A taxonomy of achieving transparent and explainable AI decisions by opening the so called black-box AI models \cite{arrieta2020explainable}. }
	\label{fig:explainbility_methods}
\end{figure*}

Explainable AI methods could be categorized, at different levels, using different criteria \cite{hall2019introduction,sokol2020explainability,arrieta2020explainable}. Figure \ref{fig:explainbility_methods} provides a taxonomy of explainable AI. There are two main categories of explainable AI, namely (i) transparent model, and (ii) Post-hoc explainability. The former represents the methods restricting the complexity of AI models for explainability while the other category represents the methods used for analyzing the models' behavior after the training. 
It is to be noted that there's a trade-off between performance (e.g., accuracy) and explanation. In literature lower accuracy has been observed for the transparent models, such as fuzzy rule-based predictors, compared to the so-called black-box methods, such as CNNs \cite{kamwa2011accuracy}. However, the explanation and intrepretability are preferred properties in critical applications, such as healthcare, smart grids, and predictive policing. Thus, there's a particular focus on developing post hoc explainable methods to keep a better balance between accuracy and transparency.

In the next subsections, we provide an analysis of how important explainable AI models are in smart city applications, and how explainable AI meets adversarial attacks and the ethical aspects of explainable AI.


\subsection{Explainable AI for Smart City Applications}
As described earlier, explainability brings several advantages to AI \cite{alonso2017building,adadi2018peeking,bellamy2018ai}. In smart city applications, its impact is more evident and crucial especially given the direct impact of the technology on society and its people. Explainable AI is particularly important in some key applications of smart cities, such as healthcare, transportation, banking, and other financial services, where key decision about humans---\textit{such as who should get a particular service? which medicine should be used? who should get a job?}---are made \cite{o2016weapons}. Such decisions in smart city applications require interpretation of data (i.e., features) to mitigate the impurities, if any, for better predictions/decisions \cite{choo2018visual}. Healthcare is one of the critical smart city applications demanding explainable AI models instead of traditional black-box AI. No doubt AI has been proven very effective in healthcare facilitating health professional in diagnosis and treatment, however, traditional black-box AI just make decisions without interpretation. Several factors are motivating the need for explainable AI in healthcare, such as the far-reaching consequences and the cost associated with a mistake in prediction \cite{ahmad2018interpretable}. Moreover, understanding the causes of AI predictions/decisions is very critical for building doctors' trust in AI-based diagnosis. Doctors would feel more confident in taking decisions given an AI-based diagnosis if the decision of the AI model is understandable/interpretable by humans. Explainable AI models would also benefit from the domain experts' knowledge to be refined. Moreover, in healthcare, the predictive performance is not enough to obtain clinical insights for decisions \cite{doshi2017towards}. In \cite{Aiwebinar,guidotti2018survey}, seven pillars of explainable AI in healthcare, showing its relationship with transparency, domain sense, consistency, parsimony, generalizability, trust/performance, and fidelity, have been provided. 

Transportation and autonomous cars is another critical smart cities application where the consequences and the cost associated with a mistake by an AI model is very high. For instance, an error in differentiating between red and green traffic lights or an error in pedestrian detection may lead to heavy losses in terms of human lives and damage to public property and vehicles. It has already happened when a self-driving Uber killed a woman in Arizona, where the object (i.e., the lady) was detected but treated it in the same way it would a plastic bag or tumbleweed carried on the wind, due to a prediction/classification error \cite{adadi2018peeking,Uber}. We believe transportation in general and autonomous cars, in particular, will benefit from explainable AI. Some interesting works, such as proposed in \cite{kim2017interpretable,soares2019explainable,haspiel2018explanations,kim2018textual}, have already been reported in the domain. 

AI models also need to be interpretable and explainable to fully explore their potential in the education sector. Despite outstanding capabilities, it is still risky to blindly follow AI models' prediction in making a critical decision in such a high-stake domain. How people will allow a machine (i.e., AI tools) to determine their child’s education? In order to trust AI in education, AI models need to make sure stakeholders (i.e., parents, teachers, and administration) understand the decision-making processes \cite{AIEducation,conati2018ai}. There are already some efforts in this directions \cite{putnam2019exploring,tulli2020explainability}. 

The literature also reports some efforts for explainable AI in defense. The concept of explainable AI has been firstly introduced in a defense project by Defence Advanced Research Project Agency (DARPA) \cite{DARPA}. Explainable AI is also a need for modern entertainment and businesses \cite{zhu2018explainable}. In order to trust in AI predictions, the prediction and decision-making process of the models should be understandable for all the stakeholders, such as investors, customers, and CEOs, etc., in the business. Table \ref{summary_explainable_ML} summarises some key explainable AI publications in different smart city applications. 

\begin{table*}[tb!]
\centering
\caption{Summary of some key works on explainable AI in terms of smart city application, type (intrinsic and post hoc), dataset, and key features of the method.}
\label{summary_explainable_ML}
\scalebox{0.78}{
\begin{tabular}{|p{.5cm}|p{3cm}|p{1.9cm}|p{3cm}|p{10.3cm}|}
\hline
\multicolumn{1}{|c|}{Ref.} & \multicolumn{1}{c|}{Application} & \multicolumn{1}{c|}{Type}& \multicolumn{1}{c|}{Dataset} & \multicolumn{1}{c|}{Description of the Method} \\ \hline
\cite{lundberg2017unified}& Healthcare & Post hoc & Self-collected & It is a game theoretic post hoc approach explaining the predictions of an AI model. To this aim, it assigns each feature an importance value based on contribution in a decision by conditioning predictions on the underlying feature. \\ \hline
\cite{lundberg2018explainable}& Healthcare& Post hoc& Self-collected& The method provides a real-time prediction of the risk of hypoxemia during a surgery. The model is trained on time-series data from a large collection of medical records. The existing explanation methods namely Model-agnostic prediction explanation \cite{bramhall2020qlime,vstrumbelj2014explaining} are employed for the explanation of the model's prediction. \\ \hline
\cite{rahman2020b5g}&Healthcare & Post hoc&Multiple online sources \cite{rahman2020b5g} & It provides a distributed deep learning based framework for COVID-19 diagnosis in a distributed environment ensuring low-latency and high-bandwidth using edge computing. Feature visualization methods are used for the explanation of the model's outcome. \\ \hline
 \cite{stirnberg2020meteorology}& Environment (air quality) & Intrinsic & SIRTA \cite{haeffelin2005sirta} & It relies on a tree-based method Gradient Boosted Regression Trees (GBRT) to predict daily total and speciated PM1 concentrations. Moreover, to further improve the performance of the model, decision trees are combined to form an ensemble prediction \\ \hline
 \cite{barredo2019lies}& Transportation & Post hoc& Madrid Open Data Portal \cite{madrid_opendata}& It relies on the existing xAI tools to extract insights from black-box traffic forecasting models namely Random Forests and Recurrent Neural Networks. \\ \hline
\cite{sun2019analyzing} & Transportation & Post hoc & self-collected& It focuses on three major steps of decision support, namely (i) synthesis of diverse traffic data, (ii) multilayered traffic demand estimation, and (iii) marginal effect analyses for transport policies. For implementation, the authors rely on the big data-driven transportation computational graph (BTCG) framework \cite{wu2018hierarchical}. The framework integrates data from several external sources including surveys, mobile phone data, floating car data etc.,. \\ \hline
 \cite{rizzo2019reinforcement} & Transportation&Post hoc & Self-collected & Proposes a reinforcement learning based solution for traffic volumes and road lanes occupancy prediction. For explanation of the outcom, the method relies on the SHAP model-agnostic technique. \\ \hline
 \cite{ghosal2018explainable} & Agriculture & Post hoc & Self-collected dataset & Proposes a deep learning framework namely xPLNet to identify and classify different types of biotic (bacterial and fungal diseases) and abiotic (chemical injury and nutrient deficiency) stresses in plant images. For better explanation, the authors rely on high-resolution feature maps isolating the visual symptoms in plants.\\ \hline
 \cite{nagasubramanian2019plant} & Agriculture & Post hoc & Self-collected & Relies a 3-D DCNN for the identification different diseases in plants in hyper-spectral imagery. The explanation purposes, the metho relies on saliency maps,visualizing the most sensitive pixels for a decision. Moreover, the method also identifies the most sensitive wavelengths used by the model for classification/differentiating in different plant diseases. \\ \hline
 \cite{wolanin2020estimating} & Agriculture&Post hoc & & It relies on a deep neural network applied to multivariate time-series of vegetation and meteorological data crop yield estimation. For the explanation of the predictions, the method makes use feature visualization techniques to analyze the relevance of the features to the predictions made by the model. \\ \hline
 \cite{reis2019explainable} & Fake News Detection &Post hoc & Buzzface \cite{santia2018buzzface} & The methods relies on Extreme Gradient Boosting (XGB) machines \cite{chen2016xgboost} for fake news detection. Explanation of the outcomes is provided using feature relevance, and observed that some features favour in detecting certain types of fake news. \\ \hline
\end{tabular}}
\end{table*}

\subsection{Explainable AI and Adversarial Attacks}
\label{explainable_adversarial_attacks}
The literature also shows a connection between adversarial attacks and explainability \cite{ignatiev2019relating,panda2018explainable,fidel2019explainability}. It is believed that explainable AI models are robust against adversarial attacks, and can help in the identification of adversarial inputs/samples by generating an anomalous explanation for the perturbed samples \cite{fidel2019explainability}. To verify the hypothesis, several efforts have been made in the literature to guard the AI model against adversarial attacks via the emerging concept of explainability. For instance, Fidel et al. \cite{fidel2019explainability} employed an explainable AI framework namely Shapley Additive Explanations (SHAP) \cite{lundberg2017unified}, which evaluates the relevance/importance of a feature by assigning it an important value for a particular prediction, to generate `\textit{XAI Signatures}' for the internal layers of a Deep Neural Network (DNN) classifier to differentiate between normal and adversarial inputs. Dhaliwal et al. \cite{dhaliwal2018gradient} proposed a gradient similarity-based approach for differentiating between normal and adversarial inputs. According to them, gradient similarity shows the influence of training data on test samples, and behaves differently for genuine and adversarial input samples, enabling the detection of various adversarial attacks with high accuracy. Some other interesting works are relying on explainable AI techniques to guard against adversarial attacks \cite{melis2020gradient,hartl2019explainability,panda2018explainable,ignatiev2019relating}. However, the explanations/information regarding the working mechanism of AI algorithms revealed by explainability methods could also be utilized to generate more effective adversarial attacks on the algorithms \cite{arrieta2020explainable}.

Adversarial AI also provides an opportunity to increase interpretability (i.e., human's understanding) of AI models \cite{marino2018adversarial,rahnama2020adversarial}. For instance, in \cite{marino2018adversarial} an adversarial AI approach is used to identify the relevance of features concerning the predictions made by an AI. The adversarial AI technique aims to find the magnitude of changes required in the features of the input samples to correctly classify a given set of misclassified samples, which is then used as an explanation of the misclassification. In \cite{rahnama2020adversarial}, on the other hand, adversarial AI techniques are used for explaining the predictions of a DNN by identifying the relevance/importance of features for the predictions based on the behavior of an adversarial attack on the DNN.

\section{Smart City and Data-related Challenges}
\label{sec:datasets}

Several challenges are associated with the collection, storage, sharing, ensuring, and maintaining the quality of data. For instance, the smart city's infrastructure requires physical resources for storing and processing the data. In addition, these resources also consume a significant amount of electricity and space as well as the environmental issues due to the carbon emissions by these resources. Smart city applications may also make use of cloud and edge deployment to overcome the lack of physical infrastructure for data storage and computing \cite{qayyum2020securing,kakderi2019smart,jan2019smartedge,qolomany2020trust}. Thanks to the recent advancement and popularity of cloud storage, the technology meets the data storage and processing requirements of a diversified set of smart city applications. However, cloud and edge deployment are also vulnerable to several adversarial, security, privacy and ethical challenges \cite{qayyum2020securing}. For instance, using third-party services may result in no control over the data, thus, the data’s privacy settings are beyond the control of the enterprise/authorities. Such deployments may also lead to potential data leakage risk by the service provider \cite{Cloud_edge,braun2018security}. Moreover, the cloud edge deployments could also be subject to several types of attacks, such as adversarial attacks, backdoor attacks, cyber kill chain–based attacks, data manipulation attacks, and Trojan attacks \cite{qayyum2020securing}.

There are also several challenges associated with the heterogeneous nature of the data, collected through several IoTs devices from different vendors, in smart city research \cite{mallapuram2017smart}. Some of the key challenges are:

\begin{itemize}
 \item \textit{Quality of the data}: The quality of the data in smart city applications largely depends on the accuracy of the IoTs devices/sensors used for collecting the data. Therefore, it should be ensured that the data infrastructure is accurate and error-free \cite{mallapuram2017smart}. In addition, some external factors, such as temperature, weather, etc., may also affect the accurate data collection. 
 \item \textit{Diversity/characteristics of the data}: Generally in typical smart city applications data is collected through several devices, making it hard to understand the characteristics of the data for removing outliers \cite{ali2016big}. Moreover, the data is collected continuously, which may result in scalability issues in the infrastructure. 
 \item \textit{Constrained Environment}: In smart city applications, generally, the devices including data collection sensors and data transfer networks have limited resources (i.e., storage, bandwidth, and processing power, etc.,) \cite{mallapuram2017smart,samie2020hierarchical}. In order to collect and transfer a large amount of data, such systems require a reliable data collection and transmission infrastructure.
\end{itemize}

In the next subsection, we will focus on some major challenges and concerns in data collection, developing, and sharing smart cities data/datasets.

\subsection{Challenges in collection and sharing Data}
The performance of AI algorithms is also constrained by the quality of the data. Thus, it is important to discuss the major challenges and issues related to dataset collection and sharing. These challenges and concerns are raised as a result of data collection, analysis, sharing, and the use of the data in sensitive applications \cite{dataethics}. The main challenges and concerns in dataset collection and sharing include informed consent in the form of understanding of how and for what purpose the data will be used, transparency, interpretation, and trust \cite{bote2019reusing,thomas2017ethical}. Though informed consent is one of the key concerns of data collection, considering the fact that future applications are sometimes unspecified and unknown, it may be inconvenient to give prior commitments regarding potential future use of the data. Moreover, data could be merged with other existing sources making informed consent even more challenging \cite{hand2018aspects}. In several cases, it is even not possible to make sure informed consent of all people subject to data collection. For instance, these days delivery by drone is very common where those who opt for free delivery consent to unlimited data collection from their home. In areas where drone delivery is permitted, a whole neighborhood could be subject to such data collection activities \cite{thippeswamy2019guide}. 

For data collection or annotation, usually, crowdsourcing studies are conducted where a large population is usually involved to collect or annotate training data for AI models in a particular application. During the process, several factors need to be considered. For instance, it is really important to inform the participants about your organization and the purpose for which the data is collected or annotated. The information of the participants should be kept confidential, and they should be allowed to withdraw from the data collection process at any time. More importantly, one should remain neutral and unbiased in conducting a crowd-sourcing study as personal preconceptions or opinions may affect the quality of the data. In the modern world, data is also collected as a result of a product/service. For instance, social media platforms can be used to collect users' data for different services. In such cases, several questions arise \cite{thippeswamy2019guide}. For instance, are the users aware of the data collection process and purpose? do they have a right and access to the data? is the company is sharing or selling the users' data? is there any policy for maintaining the informed consent if the company is sold to another one? how the companies can ensure the privacy of the users if their data is leaked to some bad actors? 

Data sharing is also subject to several questions, such as the transparency of the data, interpretation, and how much trusty the data is in a particular application? According to \cite{datasharing_ethics}, ``data sharing is not simply the sharing of data, it’s also the sharing of interpretation.'' Moreover, the re-identification of individuals or groups or linking data back to them through data mining and analysis are also key ethical concerns in data sharing. For instance, the possibility of identification or linking data to an individual or a particular group may result in gender, race and religious discrimination \cite{dataethics,taylor2016group}. Recently, a growing concern has been noticed for the transparency and interpretation of the data used for training AI models \cite{bauchner2016data,beardsley2019ethics}. For instance, Bauchner et al. \cite{bauchner2016data} emphasize the importance of data sharing in healthcare, and ethical concerns regarding data collection and sharing in the domain. Bertino et al. \cite{bertino2019data} also analyze the importance of transparency and interpretation, which they termed as providing a 360$^{\circ}$ view, of data in sensitive applications. The authors link the transparency and interpretation of data with the privacy, trust, compliance, and ethics of the data management systems.

Figure \ref{fig:data_Challenges} shows some of the major challenges in data management (collection and sharing) highlighted in the literature, and are summarized as follows:

\begin{figure}[!h]
\centering
\includegraphics[width=.99\linewidth]{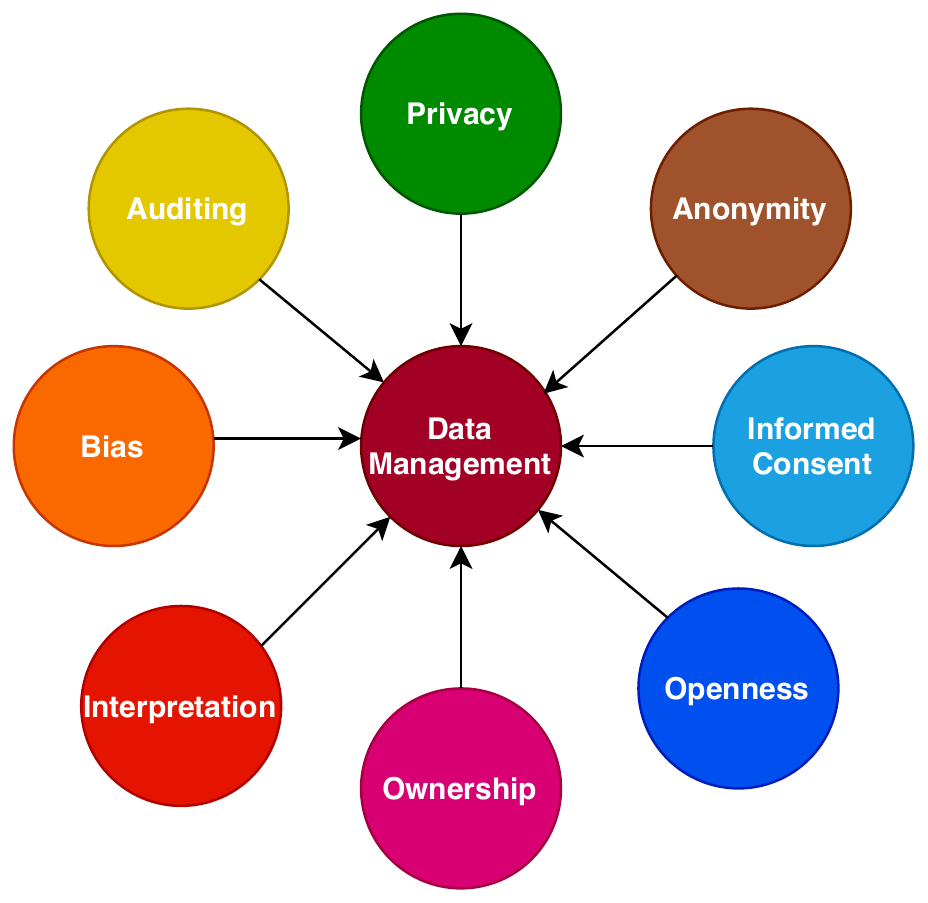}
\caption{Major challenges in data management in smart city applications.}
	\label{fig:data_Challenges}
\end{figure}
\begin{itemize}
 \item \textit{Privacy}: The biggest challenge in human-centric smart city applications is ensuring the privacy of the citizens, which is their fundamental right. 
 An improved data privacy mechanism not only helps in developing citizens' trust in different smart city services and businesses but also ensures individuals' safety as the leakage of sensitive information may endanger individuals' lives. For instance, though some off-the-shelf encryption, authentication, and anonymity techniques could reduce the chances, intelligent malicious attackers may misuse residences' sensitive information collected from smart home applications and surveillance systems to harm the individuals using a side-channel and cold boot attack \cite{li2011smart,zhang2017security,martinez2013pursuit}. Thus for the effectiveness of smart city applications, the concerned authorities should ensure that individuals' information is not misused by the authorities or any individual for any sort of personal or financial gains \cite{ali2016big}. In  recent years, there's a growing concern over citizens' privacy, and several international bodies, such as the European Union (EU), have introduced new privacy regulations. One recent example of community's concerns over privacy and radical bias is the demand for abandon of face recognition technology from  giant companies, such as Amazon, Microsoft, and IBM, for law enforcement \cite{Face_Recognition}.  To address the privacy-related concerns, various privacy-friendly techniques and algorithms have been developed using methods where AI systems’ ``sight'' is ``darkened'' via cryptography \cite{hagendorff2020ethics}. On the other hand, some believe that the traditional ``narrow'' understanding of privacy as a moral concept will eventually cease to exist and there is a need to revise the concept itself in the post-AI age \cite{hiller2016smart}. \textcolor{black}{Although it may entail some challenges, the newly introduced concept ``data philanthropy'' can also be of help in this regard \cite{lev2018data}. The basic idea that we propose here is to extend the scope of this concept to include certain cases of individuals who would voluntarily ``donate'' their data for the advancement of science or better-functioning smart cities. Within the discourse of various religious and moral traditions, there is the concept of ``charity,'' where people would voluntarily donate something they own and cherish for the benefit of others. Considering the great value that data can have in our modern world, one can argue that data would also fall within the category of valuable objects that can be donated for charitable purposes, under conditions that would vary from case to case. This will be especially applicable within the communities where familial or societal interests usually occupy a higher position than individual interests.}
 Moreover, there are also different solutions, such as differential privacy, to ensure individuals' privacy by withholding individual's information or information that could lead to identification of an individual in a dataset \cite{hassan2019differential}.  
 
 \item \textit{Informed Consent}: Informed consent, which is the process of informing and obtaining participant's consent for data collection, is a key element of data ethics. In a data collection process, it is important to make sure the users subject to data collection know about the data collection process, goal, and the way and purpose of its use in future \cite{thippeswamy2019guide}. Informed consent should fulfill four conditions including (i) the participants have information/knowledge about the data collection process, (ii) they understand the information and fully aware of the goal, future use, and the way data is collected, (iii) the participants should volunteers and should not be manipulated or persuaded in any way, and (iv) the participants should have the capabilities to understand the risks involved with the data, and able to decide whether to participate or not \cite{informed_consent}.
 \item \textit{Open Data}: For transparency and developing trust, the data and insights obtained from the data should be openly accessible. However, there are also several challenges associated with open data. For instance, it is important to determine which information should be made open, who should have an access to the data, and for what purpose the data should be allowed to be made open/used to ensure the individuals' privacy \cite{drew2016data}. 
 \item \textit{Data Ownership}: Data ownership is another key aspect of smart cities that raised serious concerns recently \cite{hummel2020own}. In smart cities, a lot of services are generally deployed by private companies, whose ultimate goal and priorities unlike public authorities are to make a profit, posing serious threats to citizens' data being monetized \cite{data_ownership}. Under these circumstance, key questions will include: who will have access to, and control over, these data? Will the upper hand be given to private companies, where the market logic will dominate or will the voice of the normal citizen count and thus more weight will be given to the public control? The answer could have been straightforward if the services were initiated and sponsored by public authorities, however, the investment from the private sector makes it very complicated. The various choices to be made in this regard will greatly determine the level of (im)morality in big data management \cite{goodman2019smart}.
 According to Ben Rossi \cite{data_nationalisation}, unfortunately in smart cities, public authorities provide the private companies with the opportunities to monetize smart city data by allowing them to deploy different services, and these companies have more information about citizens' compared to the public authorities. There are also some debates on data nationalization. For instance, Ben Rossi \cite{data_nationalisation} provides hints on how the public authorities can get hold back on the data. One of the potential solutions is to encourage joint ventures of public and private sectors where public authorities could have control over the data. Some efforts have been already noticed in this direction. For instance, the Chinese Government has initiated several joint smart city projects with big private companies. There are also debates, and some efforts have been made in terms of legislation to give citizens/users the ownership of the data \cite{user_ownership}. Moreover, there are also some solutions allowing users to retain ownership of their data while attaining different services. For instance, Bozzelli et al. \cite{bozzelli2019integrated} proposed a user data protection service. The service allows users to analyze and evaluate their data protection requirements by considering the terms and conditions of a service, which are normally overlooked by users, before using and accepting the terms and conditions of a service that might compromise their personal data. 
 \item \textit{Interpretation}: Interpretation is another key challenge of data shared and used for training AI models. For better results, the data used for training an AI model should be interpretable as also demanded by explainable AI. For instance, the big data predictive policing solution namely PredPol used by police in the USA collects and analyzes the usefulness of the data before training and making predictions about crimes in an underlying area. A very significant reduction has been observed in the crimes mainly because of the useful and interpretable data \cite{LAPD_data}. However, in smart city applications data is collected through different IoTs sensors from various vendors. Managing, interpreting, and picking relevant and useful data from such a heterogeneous and unstructured collection is a very challenging task. 
 \item \textit{Data Biases}: The datasets generally contain different types of hidden biases, either due to the collector or the respondent, in the collection phase, which are generally hard to undo and have a direct impact on the analysis \cite{latif2019caveat}. These biases are very risky in human-centric applications and need to be eliminated at the beginning. In dataset collection using surveys/questionnaires, generally two types of biases can be incorporated, namely (i) response bias, and (ii) non-response bias. The former represents the intentional bias from the respondents by giving wrong answers while the latter type of bias is encountered when no response at all is received from the respondent. One of the possible solutions for avoiding bias in such processes is to use close end surveys or restrict the respondents to some pre-defined options \cite{types_bias}. However, in smart city application data is collected from different services using different IoT sensors, and the problem of data bias is beyond the typical data collection issues. Therefore, to avoid bias in such applications a more proactive response from the citizens and authorities is needed to help in eliminating unintended bias in smart city solutions. For instance, the authorities need to invest more in research before deploying the technology in an application. Moreover, better communication and messaging strategies need to be adopted to inform and educate citizens about the goal, process, importance, and risks involved with the data collected around their city \cite{unintended_bias_cities}. 
 \item \textit{Data Auditing}: Data auditing involves the assessment of data to analyze whether the available data is suitable for a specific application or not, and the risks associated with poor data. In smart cities, data is generally collected through several IoTs sensors from various vendors, which results in an unstructured collection of data. There are several challenges associated with the unstructured collection of data as detailed earlier. Data auditing is essential, under such circumstances, to analyze and assess the quality of collected data as the performance of AI algorithms in smart city applications is also constrained by the quality of the data \cite{yu2018decentralized}. In literature, several interesting data auditing techniques have been proposed \cite{yu2018decentralized,zang2017improved,han2019lightweight,peng2020data}. For instance, Yu et al. \cite{yu2018decentralized} propose a decentralized big data auditing scheme for smart city application by employing blockchain technology. One of the key advantages of the method is the elimination of third-party auditors, which are prone to several security threats. 
\end{itemize}

Table \ref{summary_data_issues} lists some key papers on the data associated challenges in different smart city applications.

\begin{table}[]
\centering
\caption{Summary of some key works on challenges, risks, and issues associated with data collection and sharing in smart city applications in terms of application and issues covered.}
\label{summary_data_issues}
\scalebox{.8}{
\begin{tabular}{|c|c|c|}
\hline
\textbf{Ref.} & \textbf{Application} & \textbf{Challenges/Issues Discussed} \\ \hline
 \cite{patil2014big}&Healthcare & Security and privacy \\ \hline
\cite{lee2008issues} & Healthcare & Data intrepretation and fusion \\ \hline
\cite{meingast2006security}&Healthcare & Security and privacy\\ \hline
\cite{ploug2020defence}& Healthcare&Informed consent \\ \hline
\cite{swedan2020graduate}&Healthcare & Informed consent and confidentiality\\ \hline
\cite{van2020duplicitous}&Surveillance & Privacy \\ \hline
\cite{kim2020study}&Surveillance & Security and privacy\\ \hline
\cite{romanou2018necessity}& Surveillance& Privacy\\ \hline
\cite{kreuter2020collecting}&Recruitment &Privacy and informed consent \\ \hline
\cite{wachter2018normative}&Generic & Security, privacy, bias, and informed consent\\ \hline
\cite{anderson2017improving}&Generic & Informed consent\\ \hline
\cite{raghavan2020mitigating}& Recruitment& Bias\\ \hline
\cite{silberg2019notes}&Generic &Bias \\ \hline
\cite{ntoutsi2020bias}&Generic &Bias \\ \hline
\cite{roh2019survey}& Generic& Open data, intrepretation, and annotation\\ \hline
\end{tabular}}
\end{table}



\subsection{Explainability and Datasets}
\label{explainable_dataset}
 In the literature, the majority of the efforts made for explainable AI focus on the design of the algorithm to interpret AI predictions/decisions. However, other aspects are contributing to the interpretation of AI decision, such as the datasets, and post-modeling analysis \cite{chander2019creation}. For instance, a dataset used for training an AI model may contain features incomprehensible for the stakeholders, which may result in a lack of trust in the AI predictions. Therefore, to achieve better interpretation/explains of AI models' decision, explainability should be considered throughout the process starting from data/features and concluding at the post-modeling explainability \cite{chander2019creation,Explainbility_Dataset}. In this section of the paper, we focus on the explainability aspects of the dataset used for training and validation of AI models. The literature on the explainability of the dataset can be divided into four main categories, namely (i) exploratory data analysis of the dataset, (ii) description and standardization of dataset, (iii) explainable features, and (iv) dataset summarization methods. In the next subsections, we provide the details of these methods. 


\subsubsection{Exploratory analysis of the datasets}
Exploratory analysis of datasets aims to provide a summary of key characteristics, such as dimensionality, mean/average, standard deviation, and missing features, of the dataset used for training an AI model. Different data visualization tools are available to visualize different properties of a dataset and extract informative insights that could help in understanding its impact on the decisions of the AI model. For instance, Google's Facets \cite{Google_Facet}, which is an opensource data visualization library/tool, allows us to visualize and better understand data. The exploratory analysis helps in understanding the limitations of a dataset. For instance, in the case of an imbalanced dataset, such analysis could provide an early clue for the poor performance of a classifier, which can then be mitigated using different sampling techniques. 

\subsubsection{Dataset description and standardization}
AI datasets are usually released without proper documentation and description. In order to fully understand a dataset, proper description should be provided. In this regard, a standardized documentation/description of the dataset could be really helpful to mitigate the communication gap between the provider and user of a dataset. To this aim, several schemes, such as datasheets, data statements, and nutrition labels, have been proposed \cite{costa2020mt}. All the schemes aim to associate detailed and standardized documents containing a detailed description of a dataset's creation/collection process, composition, and legal/ethical considerations. Nutrition labeling, which is a diagnostic framework for datasets, provides a comprehensive overview of a dataset's ingredients helping the developers of AI models to be trained on the dataset \cite{holland2020dataset}. 

\subsubsection{Explainable Features}
Another important aspect of explainable AI is explainable feature engineering, which aims for the identification of features influencing an AI model's decision. Moreover, as one of the key characteristics of a dataset, the features should also be explainable, and make sense to the users and developers. Besides improvement in an AI model's performance, explainable features also help in the model's explainability. Explainable feature engineering can be performed in two different ways, namely (i) domain-specific feature engineering, and (ii) model-based feature engineering \cite{Explainbility_Dataset}. The former method utilizes a domain expert's knowledge in combination with the insights extracted from exploratory data analysis while the latter makes use of various mathematical models to unlock the underlying structure of a dataset \cite{shi2008daytime,murdoch2019interpretable}. For instance, Shi et al. \cite{shi2008daytime} used domain exploratory data analysis for relevant feature selection for cloud detection in satellite imagery. 

\subsubsection{Dataset summarization}
Dataset summarization is a technique to achieve a representative subset of a dataset for case-based reasoning. Case-based reasoning is an explainable modeling approach aiming to predict an underlying sample based on its similarity with training samples, which are both presented to the users for explanations. One of the main limitations of case-based reasoning is keeping track of the complete training set for comparison purposes. Dataset summarization is one of the possible solutions to avoid keeping track of the complete training set and rather selects a subset providing a condensed view of the training set.  

\section{AI Ethics and Smart Cities}
\label{sec:ethics}

AI code of ethics is another aspect of smart city applications that has recently received a lot of attention from the community. AI code of ethics is a formal document/statement from an organization that defines the scope and role of AI in human-focused applications. The three-volume Handbook of Artificial Intelligence published in 1981-1982 \cite{feigenbaum1981handbook} hardly paid any attention to ethics \cite{gebru2019oxford}. After the lapse of about three decades, the situation has radically changed. The exponential progress in the AI systems and their applications in various aspects of life have produced great benefits but have concurrently continued to trigger complex moral questions and challenges. In response, an interdisciplinary AI ethics discourse is emerging. This is owed to scholarly input from cognate disciplines, including data ethics, information ethics, robot ethics, internet ethics, machine ethics, and military ethics. These new developments were reflected in an increasing number of publications that assumed more than one form. To provide a systematic overview, relevant literature will be divided into two main categories, viz., (a) academic publications and (b) policies and guidelines.

\begin{figure*}[!h]
\centering
\includegraphics[width=.9\linewidth]{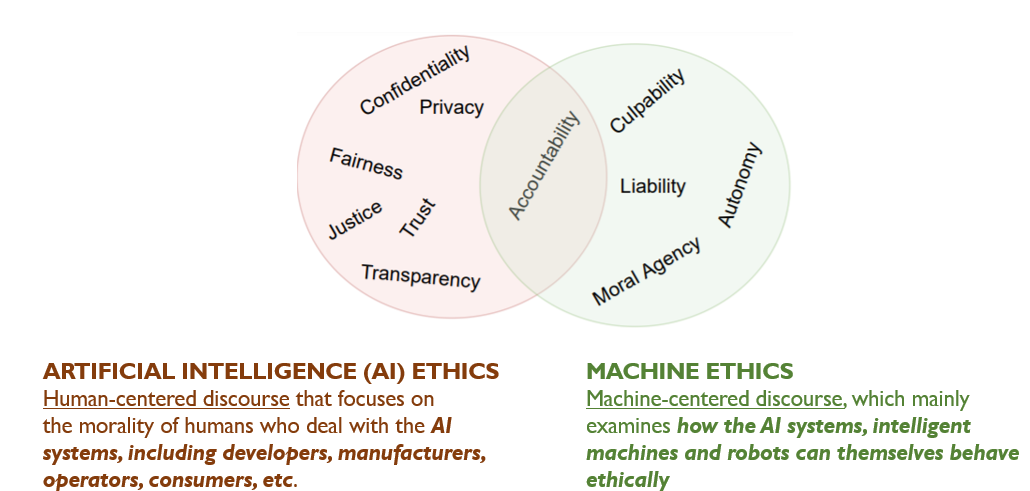}
\caption{\textcolor{black}{A visual depiction of an overlap AI Ethics (which is human centered) and Machine Ethics (which is machine centered).}}
	\label{fig:ethics_venn_diagram}
\end{figure*}

The ethical and moral discourse on the AI systems is usually divided into two main branches. The larger and more mature branch, sometimes named just ``\textit{AI ethics}'' or ``\textit{robot ethics},'' is premised on a human-centered perspective which focuses on the morality of humans who deal with the AI systems, including developers, manufacturers, operators, consumers, etc. The smaller and younger branch, usually called ``\textit{machine ethics},'' is a machine-centered discourse which mainly examines how the AI systems, intelligent machines and robots can themselves behave ethically \cite{winfield2019machine,muller2020ethics,borenstein2021emerging}. \textcolor{black}{Both the branches (i.e., human-centered and machine-centered) are overlapping as shown in Figure \ref{fig:ethics_venn_diagram}.}  We will cover review the moral questions addressed within the first branch and the second branch separately in another Sections \ref{sec:AIEthics} and \ref{sec:MachineEthics}.

\subsection{Academic Publications}
The interdisciplinary character of AI ethics was manifested in the considerably diverse backgrounds and research interests of the academics who contributed to this emerging field. Due to their close connections with AI ethics, many of the contributing authors came from the cognate fields of (moral) philosophy, engineering, and computer science. Additionally, many important authors came from other fields as well, including nanotechnology, psychology, social sciences, applied ethics, bioethics, legal studies along with some researchers who simply identified themselves as AI researchers. It is to be noted that some of the contributing authors already have an interdisciplinary background. This diverse group of researchers contributed to the AI ethics in various ways, e.g., writing book chapters, journal articles, book-length studies, editing volumes, and editing journal special issues. Below, we give representative examples of each type of these publications.

Besides individual book chapters \cite{savulescu2015moral,lagrandeur2015emotion}, important book-length studies have provided rigorous and critical insights on the moral questions of the AI systems, AI and related fields. Examples include \textit{Moral machines: Teaching robots right from wrong}, published in 2008 \cite{wallach2008moral}, \textit{Machine ethics}, published in 2011 \cite{anderson2011machine}, \textit{The Machine question: Critical perspectives on AI, robots, and ethics}, published in 2012 \cite{gunkel2012machine}, \textit{Robot ethics: The Ethical and social implications of robotics}, published in 2012 \cite{lin2012robot}, \textit{Superintelligence: Paths, dangers, strategies, published in 2014, Programming machine ethics}, published in 2016 \cite{mulgan2016superintelligence}, and \textit{Robot ethics 2.0: From autonomous cars to artificial intelligence}, published in 2017 \cite{lin2017robot}.

In addition, many individual journal articles \cite{potapov2014universal,brundage2014limitations,davis2015ethical,russell2015ethics,bench2020ethical,borenstein2021emerging,raji2021you,morley2021ethics,hickok2021lessons}, and a number of academic journals dedicated special issues to contribute to AI ethics. For instance, the \textit{Journal of Experimental \& Theoretical Artificial Intelligence} published ``\textit{Philosophical foundations of artificial intelligence},'' in 2000 \cite{akman2000introduction}, the \textit{IEEE Intelligent Systems} published ``\textit{Machine Ethics}'' in 2006 \cite{anderson2006guest}, the \textit{AI \& Society: Journal of Knowledge, Culture and Communication} published ``\textit{Ethics and artificial agents}'' in 2008 \cite{torrance2008special}, the \textit{Minds and Machines: Journal for Artificial Intelligence, Philosophy and Cognitive Science} published ``\textit{Ethics and artificial intelligence}'' in 2017, the \textit{Ethics and Information Technology} published ``\textit{Ethics in artificial intelligence}'' in 2018 \cite{dignum2018ethics}, the \textit{Proceedings of the IEEE} published ``\textit{Machine ethics: The Design and governance of ethical AI and autonomous systems}'' in 2019 \cite{winfield2019machine}, and \textit{The American Journal of Bioethics} published ``\textit{Planning for the known unknown: AI for human healthcare systems}'' \cite{chen2020planning}.

An important milestone towards the maturation and canonization of AI ethics, as a scholarly discipline, was the publication of some authoritative reference works. The \textit{Cambridge handbook of artificial intelligence}, published in 2014, included a distinct chapter on ``\textit{the ethics of artificial intelligence}'' \cite{bostrom2014ethics}. Recently, dedicated handbooks started to appear, including \textit{Handbuch Maschinenethik (handbook of machine ethics)}, published in 2019 \cite{bendel2019handbuch}, and The \textit{Oxford Handbook of Ethics of AI}, published in 2020 \cite{dubber2020oxford}, where the last chapter was dedicated to ``\textit{Smart City Ethics}'' \cite{goodman2019smart}.

These publications addressed a wide range of moral issues that are relevant to the context of smart cities, even if not explicitly stated. Thus, no serious moral discourse on smart cities can be developed without critical engagement with such publications. Additionally, an increasing number of publications started to highlight the AI moral questions within the specific context of smart cities, especially themes like privacy and information transparency. Besides the aforementioned chapter in The \textit{Oxford Handbook of Ethics of AI}, \textcolor{black}{journal articles and book chapters \cite{hiller2016smart,david2018smart,milic2018semantic,ismagilova2020security,sholla2017ethics,sholla2018docile,mark2019ethics,cardullo2019right,calvo2020ethics,offenhuber2020towards,sholla2021neuro}, one also observes a growing ethics genre with focus on smart cities. Reference works dedicated to the theme of smart cities also included chapters relevant to ethical, including The Routledge Companion to Smart Cities \cite{willis2020routledge} and the Handbook of Smart Cities, which dedicated a distinct part to ``Ethical Challenges'' \cite{augusto2019handbook}. Representative examples also include book-length studies like Data and the City \cite{kitchin2017data};  The right to the smart city \cite{cardullo2019right}; Citizens in the ‘Smart City’: Participation, Co-production, Governance \cite{cardullo2020citizens}; Technology and the city: Towards a philosophy of urban technologies \cite{nagenborg2021technology}. } 

\subsection{Policies \& Guidelines}
Besides academic researchers, AI ethics has proved to be of interest to a wide range of stakeholders. For instance, AI ethics is appealing to managers of tech giants such as Apple, Facebook, and Google, as well as politicians and policymakers. Rather than the theoretical and philosophical ramifications, which usually dominate the academic discourse, these stakeholders are more interested in applicable policies and practical guidelines that would help in developing morally-justified (self-)governance frameworks. For tech giants and multinational companies, having such policies and guidelines in hand usually serve the purpose of calming critical voices and improving the image of these companies among the general public, and particularly among their potential clients and customers. 

The efforts made by these stakeholders, especially from 2016 onwards, resulted in a great number of AI guidelines, policies, and principles. These documents and reports were surveyed, sometimes with analytical and critical insights, by some recently published papers \cite{zeng2018linking,jobin2019global,winfield2019machine,fjeld2020principled,hagendorff2020ethics}. Furthermore, some academic researchers contributed to this debate by providing theoretical foundations and critical views concerning drafting AI codes of ethics \cite{boddington2017towards,calo2017artificial}. In her work, Boddington paid special attention to the Future of Life Institute’s ``\textit{Asilomar AI principles,}'' which was the outcome of an international conference that hosted a large interdisciplinary group, with expertise in various disciplines, including law, philosophy, economics, industry, and social science \cite{boddington2017towards}. 

\textcolor{black}{From their side, almost all tech giants and multinational companies developed their own guidelines (see Table \ref{tab:tech_ethics}). After various checks, it seems that Twitter still has no published systematic AI guidelines, but this case would just represent the exception to the rule \cite{hagendorff2020ethics}.} Google has ``Artificial Intelligence at Google: Our Principles'' \cite{Google_principle} and ``Perspectives on issues in AI governance'' \cite{Google_issues}. OpenAI issued their ``OpenAI Charter'' \cite{openai2018openai}, IBM has ``Everyday ethics for artificial intelligence,'' and Microsoft has ``Microsoft AI principles'' \cite{Microsoft}. Sometimes, the adopted guidelines are the product of joint efforts and collaboration among more than one company. A good example here is the coalition ``Partnership on AI,'' where large companies like Amazon, Apple, Facebook, Google, IBM, Sony, and Intel collaborated to facilitate and support the responsible use of AI \cite{hagendorff2020ethics}. Table \ref{tab:tech_ethics} provides an overview of moral principles in the AI codes of Tech companies. One recent example of considering the ethical aspects of AI in human-centric applications from these companies is quitting the use of face recognition technology for law enforcement after the privacy and racial concerns over it from the community \cite{Face_Recognition}.    

At the governmental level, many countries drafted guidelines and policy frameworks for AI governance. The two leading AI superpowers, China and the United States, were at the forefront in this regard. For the USA, there are several documents and reports, including the ``Preparing for the future of artificial intelligence'' published in 2016 and ``The National artificial intelligence research and development strategic plan: 2019 update,'' by the National Science and Technology \cite{parker2018creation,bundy2017preparing,national2019national}. As for China, there is the ``Beijing AI Principles'' issued in 2019 by the Beijing Academy of Artificial Intelligence and backed by the Chinese Ministry of Science and Technology \cite{beijing2019principles}.

\textcolor{black}{At the transnational or global level, there are also important initiatives \cite{ebell2021towards,schiff2021ai}. The Institute of Electrical and Electronics Engineers (IEEE) produced two versions of the ``Ethically Aligned Design.'' The first version came out in 2016 and the second in 2019 \cite{chatila2019ieee,shahriari2017ieee}. After open consultation on a draft made publicly available in December 2018, the European Commission published ``Ethics guidelines for trustworthy AI'' \cite{Pekka}. The last example to be mentioned here is the intergovernmental Organization for Economic Co-operation and Development (OECD), which adopted in May 2019 the ``OECD Principles on AI.'' The document is meant to promote innovative and trustworthy AI that respects human rights and democratic values \cite{OECD}. Inspired by this initiative, the G20 adopted the human-centered AI Principles \cite{OECD,G20AI}.} 

\begin{table}[!h]
\centering
\caption{Overview of moral principles in the AI codes of Tech companies
.}
\label{tab:tech_ethics}
\scalebox{.8}{
\begin{tabular}{|p{1.5 cm}|p{7 cm}|}
\hline
\textbf{Company} & \textbf{Key Principles} \\ \hline

 
IBM & \vspace{-1.0\baselineskip} \begin{itemize}
 \item Accountability
 \item Value alignment
 \item Explanability
 \item Fairness
 \item User data rights
\vspace{-1.0\baselineskip} \end{itemize} \\ \hline
 
 Google & \vspace{-1.0\baselineskip} \begin{itemize}
 \item No overall harm to society
 \item No use of AI for weapons
 \item No violation of human rights and international law
\vspace{-1.0\baselineskip} \end{itemize} \\ \hline
 
 Microsoft & \vspace{-1.0\baselineskip} \begin{itemize}
 \item Fairness
 \item Security and privacy
 \item Empower and engage everyone
 \item Transparency and interpretability
 \item Accountability 
\vspace{-1.0\baselineskip} \end{itemize} \\ \hline
 
Samsung & \vspace{-1.0\baselineskip}\begin{itemize}
 \item Equality and diversity
 \item No unfair bias
 \item Easy access for all
 \item Explainability
 \item Social and ethical responsibility
 \item Benefit to society and corporate citizenship
\vspace{-1.0\baselineskip} \end{itemize} \\ \hline

Intel & \vspace{-1.0\baselineskip}\begin{itemize}
 \item New employment opportunities 
 \item People’s welfare
 \item Accountability and responsibility 
 \item Privacy 
\vspace{-1.0\baselineskip} \end{itemize} \\ \hline
\textcolor{black}{Facebook} & \vspace{-1.0\baselineskip}\begin{itemize}
 \item \textcolor{black}{Privacy \& Security} 
 \item \textcolor{black}{Fairness \& Inclusion}
 \item \textcolor{black}{Robustness \& Safety}
 \item \textcolor{black}{Transparency \& Control}
 \item \textcolor{black}{Accountability \& Governance}
\vspace{-1.0\baselineskip} \end{itemize} \\ \hline
 \textcolor{black}{Global convergence of AI code of ethics} & \vspace{-1.0\baselineskip}\begin{itemize}
  \item \textcolor{black}{Transparency}
  \item \textcolor{black}{Fairness}
  \item \textcolor{black}{Non-maleficence}
   \item \textcolor{black}{Responsibility}
   \item \textcolor{black}{Privacy}
\vspace{-1.0\baselineskip} \end{itemize} \\ \hline
\end{tabular}}
\end{table}
\subsection{Analytical review of the key issues}
\begin{figure}[!h]
\centering
\includegraphics[width=1.05\linewidth]{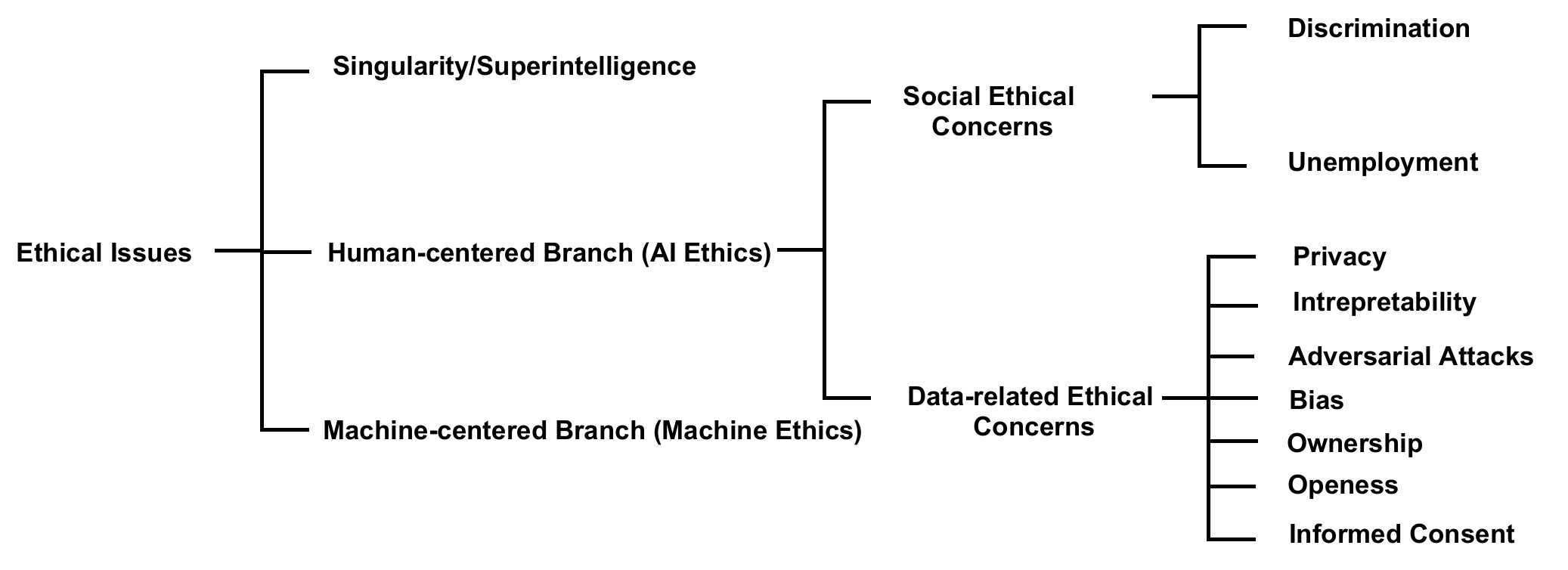}
\caption{A taxonomy of ethical issues.}
	\label{fig:ethics_taxonomy}
\end{figure}

Figure \ref{fig:ethics_taxonomy} provides a taxonomy of the key ethical issues discussed in the literature. In this section, we will mainly focus on the algorithmic issues as a detailed description of the data ethics has been provided in Section \ref{sec:datasets}.

Before delving into the detailed issues addressed by the above-sketched literature (see overview below and in Table \ref{tab:ethical_issues}), a methodological note is in order. Due to the popularity of AI and the polarizing debates in media, some of the contributors to the field of AI ethics stresses the need to distinguish between genuine and pretentious moral problems and to stress that this field should focus on the former rather than the latter type of problems \cite{muller2020ethics,powersethics}. 
\begin{table*}[!h]
\centering
\caption{Overview of the key issues that (should) deserve attention in the moral discourse on AI. The significance of some issues is agreed upon (Serious Issues), where other issues are viewed as less important or simply non-issues (Pretentious Issues). The (in)significance of some other issues, mainly represented by the singularity hypothesis, is still a point of controversy and agreement.}
\label{tab:ethical_issues}
\scalebox{.7}{
\begin{tabular}{|c|c|c|c|c|}
\hline
\textbf{Pretentious Issues} & \multicolumn{4}{c|}{\textbf{Serious issues}} \\ \hline
\begin{tabular}[c]{@{}c@{}}Worries raised by\\ the Hiroshi Ishiguro’s \\ Geminoid\end{tabular} & \multicolumn{3}{c|}{\textbf{\begin{tabular}[c]{@{}c@{}}(A)\\ Human-centered branch (AI Ethics)\end{tabular}}} & \textbf{\begin{tabular}[c]{@{}c@{}}(B)\\ Machine-centered branch (Machine ethics)\end{tabular}} \\ \hline
\multirow{3}{*}{\begin{tabular}[c]{@{}c@{}}Worries raised by \\ the Humanoid robot Sophia, \\ especially when she delivered \\ a talk at the UN\end{tabular}} & \multicolumn{2}{c|}{\begin{tabular}[c]{@{}c@{}}(A.1)\\ Data-related concerns\end{tabular}} & \begin{tabular}[c]{@{}c@{}}(A.2) \\ Social concerns\end{tabular} & \multirow{3}{*}{\begin{tabular}[c]{@{}c@{}}Accountability\\ Autonomy\\ Culpability\\ Liability\\ Moral agency\end{tabular}} \\ \cline{2-4}
 & Preserving key values & \begin{tabular}[c]{@{}c@{}}Confidentiality\\ Privacy\\ Accountability\\ Fairness\\ Justice\\ Transparency\\ Trust\end{tabular} & \multirow{2}{*}{\begin{tabular}[c]{@{}c@{}}Discrimination\\ Undermining Inter-human relations\\ Unemployment\end{tabular}} & \\ \cline{2-3}
 & \begin{tabular}[c]{@{}c@{}}Addressing the moral implications \\ of technical problems\end{tabular} & \begin{tabular}[c]{@{}c@{}}Safety \\ Adversarial attacks\\ Bias\\ Explainability\end{tabular} & & \\ \hline
\multicolumn{5}{|c|}{Singularity hypothesis (?)} \\ \hline 
\end{tabular}}
\end{table*}
The publicity of certain ``exotic'' anecdotes and their wide circulation in media would make people mistakenly think that they raise genuine ethical issues. This holds true for the public unveiling of the Japanese roboticist Hiroshi Ishiguro’s Geminoids, an android that so closely resembles his own appearance and does human-like movements, such as blinking and fidgeting with its hands. Another example here is the robot ``Sophia,'' which received the ``citizenship'' status from Saudi Arabia after her speech at a United Nations meeting \cite{powersethics}. It is to be noted that it is quite difficult to get the Saudi citizenship, even for people who were born in this country and spent a great deal of their life there but had no Saudi parents. Such incidents make some people imagine or create fearful scenarios that ethicists and policymakers should urgently address their moral ramifications, as if they are part of an already existing dilemma. However, Ishiguro’s robot is a remotely controlled android, not an autonomous agent and the speech given by Sophia was not her own work but it was prerecorded by an organic human female. Thus, the fears and concerns promoted after such incidents are more pretentious in nature and are usually viewed as non-issues, from a moral perspective. They come close to analogous claims made about earlier technologies, e.g., writing will destroy memory, trains are too fast for souls, telephones will destroy personal communication, video cassettes will make going out redundant, etc. \cite{muller2020ethics}.

Moral philosophers argue that such ``non-issues'' should not be part of the mainstream AI ethics \cite{shanahan2015technological,powersethics,muller2020ethics}. However, sometimes it proves difficult to agree whether some AI-related questions and challenges should be considered as genuine or pretentious issues. The main example here is the so-called ``singularity hypothesis'', which will be discussed in a distinct section below.

\subsubsection{\textbf{Singularity/Superintelligence}}
Unlike the usual concern linked with most technological advances, viz., undermining people’s health or wellbeing, advances in the AI systems (sometimes together with the related field of neurology) is believed to pose an existential threat to the human species altogether. This concern is usually couched under the so-called ``singularity hypothesis''. 

The basic idea of this hypothesis is that once the AI systems are able to produce machines or robots with a human level of intelligence, these machines will also be able to act autonomously and create their own ``superintelligent'' machines that will eventually surpass the human level of intelligence. With such a shift-making sequence of developments, the point of ``singularity,'' similar to that of physics, will be a natural outcome. After this point, the superintelligent machine will be the last invention made by man because humans will not be able to have things under control anymore, including their own destiny. Consequently, human affairs and basic values in life (including even what it means to be human), as we understand them today, will collapse \cite{shanahan2015technological,muller2020ethics}. 

For those who believe in the singularity hypothesis, one of the possible post-singularity scenarios is that humans will be replaced by superintelligent machines and thus mankind will become obsolete. The proponents of a more optimistic scenario do not speak of human extinction but of transformation into superhuman intelligent beings. Owing to mutual hybridization between men and machines, humans will be able to exponentially increase their levels of intelligence, all other capacities, and their lifespan up to the possibility of achieving immortality \cite{shanahan2015technological,muller2020ethics}. 

On the other hand, some voices consider the singularity hypothesis dubious, untenable and overestimation of AI risks. Thus, some wonder whether this hypothesis ever deserves to be viewed as a real moral issue or it should actually be seen as something imaginary whose right place is science fiction rather than moral discourse \cite{shanahan2015technological,powersethics,muller2020ethics}. The critics of the singularity hypothesis sometimes even accuse its proponents of lacking work experience in the AI field \cite{calo2017artificial,hagendorff2020ethics}. Such reservations about the singularity hypothesis and questioning whether it is even a serious issue to be addressed may explain the silence of many of the above-reviewed policies and guidelines on this issue 
\cite{national2019national}. Even the report released in 2017 by the US Center for a New American Security (CNAS), which had the term singularity in its title, did not provide serious analysis of the singularity hypothesis \cite{kania2017battlefield}. When the aforementioned ``Preparing for the future of artificial intelligence'' specifically touched upon the singularity hypothesis, it was stated that it should have little impact on current policy and that it should not be the main driver of AI public policy \cite{national2019national}. The same attitude was adopted by the first version of the IEEE’s ``Ethically aligned design,'' where an implicit reference was made to the singularity hypothesis warning against adopting ``dystopian assumptions concerning autonomous machines threatening human autonomy'' \cite{chatila2019ieee}. 

\subsubsection{\textbf{Human-centered Branch (AI Ethics)}}
\label{sec:AIEthics}

\paragraph{\textit{Data-related concerns (e.g., privacy, transparency, explainability, adversarial attacks)}}

Broadly speaking, the efficiency of AI systems heavily depends on the quality of the training data. Thus, a great deal of the AI moral issues and dilemmas revolve around the central question of how such big data should be managed in an ethical way. While trying to collect and process as much data as possible, the AI systems can actually be seen as performing a modernized form of the conventional state surveillance by secret services. Various techniques that can be used in smart cities, such as face recognition and device fingerprinting, in combination with ``smart'' phones and TVs, ``smart governance'' and ``Internet of Things,'' are tools for huge data-gathering machinery. As some observers stated, the resulting data will not only include ``private'' or ``confidential'' information about us but these tools will even know more about us than what we know about ourselves. Consequently, the data gathered can be used to manipulate one’s behavior. Besides the possibility of deploying it to infringe upon people’s privacy and confidentiality of information, this massive data-gathering machinery can also make money through our collected data without consenting or even informing us. This is sometimes called ``surveillance economy'' and ``surveillance capitalism'' \cite{burr2019can,zuboff2019age,muller2020ethics}. A more detailed discussion on the data-related ethical issues and concerns, such as privacy, bias, ownership, data openness, interpretation, and informed consent, has been provided in Section \ref{sec:datasets}.



Explainability---which is closely related to key moral concepts such as fairness, bias, accountability, and trust---is another significant aspect of big data management. The minimum level of required explainability intersects with the concept of transparency, which would simply mean developing easily-understood overview of system functionality. In other words, the AI systems should at least maintain precise accounts of when, how, by whom, and with what motivation these systems have been constructed, and these accounts should be explainable and understandable. Moreover, the very tools used to build the AI systems can be set to capture and store such information \cite{bryson2019artificial}. On the other hand, explainability, as a technical term, has further moral requirements. It means that the causes behind an AI model’s decision should be explainable and understandable for humans so that stakeholders can be aware of the AI model’s biases, potential causes of the bias, etc. \cite{sado2020explainable}. The lack of explainability and transparency, which will be seen as opacity, continues to trigger public and scholarly debates about the possible moral violations related to discrimination, manipulation, bias, injustice, etc. An AI algorithm developed by Goldman Sachs was said to be discriminating against women \cite{Applecard}. Also, the Google Health study, published in \textit{Nature}, which argued that an AI system can outperform radiologists at predicting cancer, was said to violate transparency and reproducibility \cite{mckinney2020international,mckinney2020reply,haibe2020transparency}.
To address such concerns, the AI field has been developing techniques to facilitate the so-called ``explainable AI'' and ``discrimination aware data mining'' \cite{hagendorff2020ethics}. On the other hand, governmental efforts continue to put pressure on the AI industry to produce more explainable applications. For instance, the EU General Data Protection Regulation (GDPR) underlined the ``right to explanations'' \cite{sado2020explainable}. Furthermore, the aforementioned EU ``Ethics guidelines for trustworthy AI'' included the principle of explicability, as one of the four core ethical principles in the context of AI systems \cite{Pekka}.

Another major concern related to data governance has to do with ensuring its security and developing protective measures against adversarial attacks, which can have a serious impact on the AI systems. AI algorithms, whose behavior fairly shapes life in smart cities, mainly feed on data collected from every participated device to have fully integrated complex smart solutions. However, AI algorithms are not safe by nature since the adversarial attacks have been approved in different smart domains. This creates deep ethical responsibility shared by all stakeholders to ensure data safety for both assets and people, to the extent that some considered it a human-rights issue \cite{yeung2019ai}. 

Several defense techniques have being developed to mitigate or minimize the risk of adversarial attacks  \cite{samangouei2018defense,zantedeschi2017efficient,zhang2019defense}. Also, the Generative Adversarial Networks (GANs) support decision systems in several smart areas by generating realistic examples to enrich the available data set (data augmentation) and thus improve the efficiency of the AI models \cite{shao2019generative,antoniou2017data,gao2020data}. It is to be noted that the commissioned cyberattacks, originally meant to test the immunity of the AI systems to the threat of Adversarial AI or Offensive AI, can also help address some of the aforementioned concerns. For instance, it can step in to secure fairness in AI solutions so that classifiers will not judge based on any protected attributes related to gender, religion, rich, poor, etc. \cite{delobelle2020ethical} and this is called adversarial fairness. They may also need to keep a level of privacy of some sensitive data and this is called adversarial representation \cite{martinsson2020adversarial,ericsson2020adversarial}. Such benefits explain the presence of a clear trend in literature to expose all possible adversarial attacks on different systems. This can be viewed as part of typical ethical hacking, where AI specialists look for every possible form of attack to improve the process of defenders’ development.

\paragraph{\textit{Social concerns (e.g., discrimination, unemployment)}}

In addition to the problems highlighted above, big data misgovernance can also create social problems. For instance, the absence of explainability can pose a serious threat to democracy; the so-called ``threat of algocracy'' \cite{danaher2016threat}. This threat will likely happen by standardizing dependence on ``intelligent'' systems whose rationale or mode of reasoning for the decision they made is inaccessible to individual citizens and sometimes even experts \cite{muller2020ethics}. 

Additionally, the aforementioned example of Goldman Sachs and similar stories show that data-driven algorithms can contribute to sexism, racism, or reproducing other negative stereotypes that we collectively agreed to judge as bad, even if they sometimes reflect part of our current reality. Unregulated usage of AI applications like automated facial analysis proved to have systematic biases by skin type and gender \cite{gebru2019oxford,bryson2019artificial}. Instead of helping us reform the exiting inequalities in societies, mathematical models and algorithms often reinforce them \cite{o2016weapons}. To address such biases and discriminatory stereotyping, more carefully programmed AI systems are being developed. For instance, some discrimination-sensitive programs can be used in the early stages of human resources processes to help shortlist diverse CVs \cite{bryson2019artificial}. What is also important in this regard is that the AI field itself should be more inclusive and diverse when it comes to the cultural and ethnic background and gender of the AI teams \cite{dignum2018ethics}. 

By its increasing ability to outsource skilled and unskilled jobs, another socio-economic concern is that AI will disrupt the labor market. The pessimistic view sometimes goes as far as to warn of a dystopian climax, where a handful of AI giants will take jobs away from millions of people who will end up having nothing to do except ``entertaining'' themselves by what the AI industry would allow them to access. At the other end of the spectrum, there is an optimistic view whose advocates promise of an AI utopia where the AI systems will generate wealth, create more jobs, and improve the overall economic growth. One of the key challenges to properly navigate these concerns is that there is little economics research in this area and available predictions are premised on past technologies. This state of academic research makes it difficult for policymakers to prepare well for the prospective AI impact on the labor market and economy in general \cite{goldfarb2019economics,boddingtonnormative,moradi2020future}. Beyond the AI positive or negative economic impact, some researchers expressed specific concerns in certain applications, like the so-called ``carebots,'' which are meant to offload caregiving to a machine. Even if this automation of caregiving will not result in job cuts, replacing human care will still have social costs, e.g., exchanging feelings and emotions among humans will cease to be part of caregiving \cite{donathethical}. 

\subsubsection{\textbf{Machine-centered Branch ( Machine Ethics)}}
\label{sec:MachineEthics}

The machine-centered branch of AI ethics, or ``machine ethics,'' approaches machines as subjects or agents, rather than objects or tools used by humans. Despite some vagueness about the exact scope and subject of this branch, the basic idea is that ``machine ethics'' discourse would focus on questions related to the morality of the machine itself, e.g., can a machine behave ethically towards humans or other machines? and if yes, which moral standards should apply to judge this behavior? Would the machine in such a case be held accountable, morally responsible, or holder of rights and obligations? \cite{winfield2019machine,muller2020ethics}.

Available research shows a variety of approaches, already applied in experimental demonstrations with robots, that explore how the machine can be trained to recognize and correctly respond to morally challenging situations. It is to be noted that the outcome of these trials is still far away from producing even a human-like being whose acts can be judged in the same way we judge human moral agents. Researchers just speak about ``robots with very limited ethics in constrained laboratory settings'' \cite{winfield2019machine}. In order to accommodate the restricted moral autonomy in some (future) AI systems, some researchers proposed multi-layered typologies for ethical agents. In these typologies, the highest category of full ethical agents is (now) exclusive to an average adult human whereas the machines trained to behave ethically fall under lower categories \cite{moor2006nature}.

Whatever one’s conviction is about the nature of morality that can be assigned to certain AI systems and how far we can regard them as ``artificial moral agents,'' the very idea itself raised complex questions about key concepts like moral responsibility, accountability, and liability. This holds particularly true for the two famous AI applications, namely autonomous vehicles and autonomous weapons. In principle, such applications challenge the conventional idea that whenever there is a victim, there should an identifiable culprit. The victims of violations made by autonomous cars or weapons will face the difficulty of allocating punishment, sometimes called the ``retribution gap''. This is because they will not have a human driver or shooter who can be held accountable \cite{danaher2016robots,muller2020ethics}. In response to these difficulties, proposals were made to forgo the idea of accountability assigned to a specific individual (e.g., the motorist or the shooter) and to assign it to a pool of involved stakeholders (e.g., programmers, manufacturers, and operators of the AI systems, besides the bodies responsible for taking infrastructure, policy and legal decisions, etc.) \cite{anderson2011machine,muller2020ethics,kroll2020accountability}.

\section{Insights and Lessons Learned}
\label{sec:insights}
In this section, we present the insights and lessons learned from the literature on each challenge to AI in smart cities. 

\subsection{Smart City AI Security and Robustness}
The topic of adversarial AI is not an emerging topic, however, it becomes a crucial and hot topic in the era of smart cities which needs extra efforts to reach an acceptable level of trust in our AI-based products. Trust might be defined concerning the possible attacks, defense mechanisms, and the expected effect on the overall system. This may create a trade-off between safety and performance which needs further exploration. 

\textcolor{black}{The AI safety strategies come in four categories based on four general safety strategies in engineering \cite{varshney2017safety}. We highlight the basics of each of them with possible examples related to the discussion in this section. }
\begin{itemize}
\item \textbf{\textcolor{black}{Safe Design Strategy}}: \textcolor{black}{The main idea in this strategy is to study the data and any potential bias or harm before building AI solutions.  For example, training a model on a mix of animals and humans could lead to harmful results. Using a dataset that is biased to specific classes such as lighter-skin examples are overwhelming in a dataset compared to other darker-skin colors could also be a biased solution towards specific classes \cite{ buolamwini2018gender}. Therefore, the imbalance of the examples in the dataset forces the classifiers to perform better, in terms of accuracy, with specific classes related to male over female, and lighter-skin color over darker-skin color. The general purpose IBM face recognition was stopped because it was used for racial profiling, the MIT technology review showed that this software does well with lighter-skined color female than dark-skin color female\footnote{https://www.technologyreview.com/2020/06/12/1003482/amazon-stopped-selling-police-face-recognition-fight/}.} 
\item \textbf{\textcolor{black}{Safety reserves}}: \textcolor{black}{The feature set could be partitioned into protected, such as gender, race…etc, and unprotected groups where the risk ratio of harm of a protected group to an unprotected group should not exceed a  predefined threshold.}
\item \textbf{\textcolor{black}{Safe Fail}}: \textcolor{black}{If the decision cannot be given with confidence, the rejection option would be the choice. The human would step in to have manual decisions. }
\item \textbf{\textcolor{black}{Procedural Safeguards}}: \textcolor{black}{The availability of the open-source machine learning algorithms could improve the testing and auditing works. However, since the data is playing a major role in any AI-based solution, the open dataset; freely available, could help in developing more safe applications.}
\end{itemize}

\textcolor{black}{Although the above strategies could improve the safety of AI-based solutions, several defense methods have been developed against security attacks to maintain the safety of AI-based applications. }

Some key lessons learned from this section are summarized
as:
\begin{itemize}
 \item Adversarial attacks are proved in several smart city applications and they have serious consequences on people's lives, privacy, opportunities, and assets. They could also significantly impact the economy and the environment of countries.
 \item All stakeholders in developing smart city applications are ethically responsible to follow the good technical practices and extensively evaluate the impact of any AI applications on fairness, privacy, and lives. 
 \item Anti-adversarial attack solutions are not magic and all authorities and organizations share the responsibility of risk prevention and mitigation.
 \item Adversarial data do not mean ``harm'' all the time, it can be utilized as a data augmentation technique and to build more robust AI-based solutions.
 \item Due to their high severity, adversarial attacks should be put into the educational track as an integral part of the model building and deployment process of AI applications.
 \item The transferability of adversarial examples across models enables the attacker to target even the black-box model. There is no effective defense mechanism currently exist which shed the light on the techniques of substitution models. 
 \item Organizations may need to invest more not only in their collected data but also in securing the models they developed. This probably needs more budget on security, training, and tools.
 \item AI models that show high accuracy at testing time could not be good choices if the robustness of the model against attacks becomes part of the evaluation process.  
\end{itemize}

\subsection{Smart City AI Interpretability}
Despite the outstanding capabilities, the decisions/predictions made by the traditional black-box AI algorithms are not straightforward, in fact un-understandable, for different stakes-holders, such as government authorities and citizens, involved in a smart city application. Even the data scientists that created the model may have trouble explaining why their algorithm made a particular decision. One way to achieve better model transparency is to adopt from a specific family of models that are considered explainable. Even, sometimes the developers of the AI models are not fully aware of the causes of a particular decision. Understanding the causes of a model's decision, in general, and in smart city applications in particular, are critical for developing users' trust in the system. For instance, in healthcare, understanding the causes of AI predictions/decisions is very critical for doctors to consider AI-based clinical insights. Doctors would feel more confident in taking decisions given AI-based diagnosis if the decision of the AI model is understandable/interpretable by a human. Explainability also provides an opportunity for AI models/developers to benefit from the domain experts' knowledge to deal with the impurities in data and structure of the models. 

Some key lessons learned from this section are summarized as:

\begin{itemize}
 \item A lot of interest and demand has been observed for explainable AI over the last few years.
 \item Explianibility helps in building stakeholders' trust in AI models' predictions, which will ultimately speed up its adoption in critical smart city applications, such as healthcare. 
 \item Explainability also plays a vital role in ensuring fair AI decision by identifying and eliminating decisions based on protected attributes such as race, gender, and age.
 \item There's a trade-off between explanation and performance. Transparent models are good for explanation, however, their performance is lower compared to the black box models, such as deep learning models.
 \item There's a deep connection between explainability and other emerging concepts in AI, namely adversarial attacks and ethics.
 \item Explainability helps AI models to guard against adversarial attacks by differentiating between genuine samples and adversaries. 
 \item Explainability and ethics also link and cross-fertilize each other in AI.
\end{itemize}

\subsection{AI Ethics}
The literature reviewed in this section demonstrates a growing interest and concern over the ethical aspects of the AI systems and their applications. A diverse group contributed to the emerging field of AI ethics, including not only academics and researchers but also governments and tech giants, such as Apple, Facebook, and Google. They all have realized the growing impact of AI technology on society and believe that ethical deliberations, guidelines, and governing policies are necessary to make a rigorous trade-off between potential benefits and possible harms. 

The key lessons learned can be summarized as follows:

\begin{itemize}
\item AI ethics is increasingly moving towards a distinct scholarly field of inquiry with strong interdisciplinary character. Besides the two main involved groups, namely philosophers and engineers, this young field is also benefiting from insights provided by an interdisciplinary group of scholars, researchers, and practitioners.
\item In their attempt to canonize the young field of AI ethics and to theorize and standardize its scope, questions, and methodology, various academic journals and publishers have been actively producing books, edited volumes, journal special issues, and recently also handbooks. 
\item The key players in the AI industry, including multinational companies alongside national and transnational governmental bodies, drafted various policies and guidelines meant to demonstrate their commitment to ethical governance of their activities in the AI industry.
\item The wide range of moral issues addressed by academic publications and/or guidelines show disagreement on certain issues (such as the singularity hypothesis) and whether they should be regarded as real problems. On the other hand, a great number of issues were consensually viewed as serious challenges, including those with relevance to smart city applications. Representative examples were discussed under broad themes, including big data management (e.g., privacy, explainability, transparency, opacity, bias), social problems (e.g., facilitating discrimination and disrupting the labor market). 

\end{itemize}

\section{Open Issues and Future Research Directions}
\label{sec:research_directions}
\subsection{Smart City AI Security and Robustness}
Google scholar shows growth in the number and scope of adversarial attacks research since the last decade \cite{qayyum2020securing}. The collaboration of multidisciplinary teams including data scientists, cybersecurity engineers, and domain-specific-professionals is needed for adversarial attacks research and development. Future research is expected to set a policy to accurately describe ethical outlines, and how and when the AI should be part of the organization's ecosystems\cite{finlayson2018adversarial, finlayson2019adversarial}. Some of possible research opportunities and open issues are:

\begin{itemize}
\item \emph{Performance and Accuracy vs. Security.}

The classical trade-off between the response time and the safety procedure would be the first concern raised in deploying AI in smart cities where decisions are supposed to be taken on time. Applying detection algorithms against adversarial attacks must be carefully evaluated in different fields especially those that depend mainly on fast decisions such as autonomous vehicles (AVs). Another concern related to performance is the accuracy of AI models when these are trained on both benign and adversarial data, i.e., the false positive and true negative rates. Different parameter optimization methods of learning-based algorithms share the same objective, i.e., maximizing the overall accuracy of the model\cite{qolomany2017parameters}. However, the interesting question would be: do those parameters have any impact on the model’s immune system against adversarial attacks?

\item \emph{Estimating attacks implications (The ripple effect).} 

The ripple effect of the attacks must be considered in future works. Given the complexity of the smart city's ecosystems, attacking one model may have a series of consequences on the whole city and also may unintentionally attack other models. Estimating the loss and effect of attacking a model and functional dependency evaluation could be integral parts of the future AI-based systems development life-cycle. We can expect more interest in simulation works in this area soon. 

\item \emph{Real-Time Adversarial Attacks}.

This is another challenge for AI safety teams. There is a need to evaluate the current techniques in generating poisoning data when only part of benign data is available, i.e., streaming. How about the structures of defenders in real-time environments?\cite{ gong2019real}

\item Future works may show more efforts in defining the rules on operating smart cyber-systems and the accountability of the services providers and operators\cite{ lim2019algorithmic}. 

\item \emph{Unintentional attacks in smart waste and agriculture}
Smart waste and agriculture mainly depend on a network of sensors that work in harder conditions compared to some other fields such as transportation. In such a scenario, the environments might be wet, humid, dirty, have different temperatures, and may suffer from pollution. For example, the sensors attached to the animals in large farms, sensors on trash bins, electrochemical sensors for soil nutrients, etc, are subject to convey some noise besides the required data due to the environmental effects. This could be an important source of unintentional attacks that should be evaluated and taken into account in future works. 

\item \emph{AI models detection and isolation techniques}
In \cite{khanapuri2019learning}, a technique of abnormal vehicle behavior detection and isolation is applied on the object level (i.e., vehicle)  which may run several models to control the driving tasks and traffic management. Evaluating the approach on a lower level, i.e., models, to detect and isolate the possibly attacked models might add value to the overall safety. Developing guidelines for replacing suspected models or defining alternatives in AI models' maintenance plan could improve consumers' trust.

\item \emph{Robustness and safety metrics to be involved in the evaluation process of AI models.} The current metrics to evaluate the performance of AI models could take into account the factor of safety and the robustness of the model against different types of attacks. AI models of high accuracy at testing time might be the worse with a little noise added by attackers at production environment\cite{biggio2018wild}. This leads to the possible need of revisiting the AI models evaluation policy before deployment. Thus, the agreement between the stakeholders or services. 

\end{itemize}

\subsection{Smart City AI Interpretability}
Although a lot of efforts have been made for the interpretation/explainability of AI algorithms since the concept of explainable AI has been introduced. However, there are still many aspects of explainable AI that need to be analyzed. In this section, we provide some of the open issues and future research directions in the domain.

\subsubsection{Interpretation vs. Performance}
Despite all the benefits it brings for all the stakeholders in different application domains, there are some concerns about its impact on the performance and the development process. It is believed that the efforts for explainability will not only slow down the development process but also put constraints on it, which might also hurt the performance (i.e., accuracy) of the models \cite{gilpin2018explaining}. For better interpretability AI models to be simple as simpler the model more explainable is the causes of an underlying decision. However, literature shows that usually complex AI algorithms (e.g., deep learning) tend to be more accurate. The trade-off between explainability and performance is believed to be optimized with better explainability methods, which is one of the key research challenges in the domain \cite{arrieta2020explainable,gunning2017explainable,arrieta2020explainable}. 

\subsubsection{Concepts and Evaluation Metrics}
The literature still lacks a common ground, structure, and a unified concept of explainability \cite{arrieta2020explainable}. However, several efforts have been made in this regard. For instance, Arrieta et al. \cite{arrieta2020explainable} attempted to provide a common ground or a reference point in this regard. According to them, the explainability of an AI model refers to its ability to make its functioning (i.e., causes of its decisions) clearer to an audience. The authors also emphasize the need and definition of an evaluation metric or set of metrics for the evaluation and comparison of AI models in terms of explainability and interpretation capabilities. 

\subsubsection{Explanation of Deep Learning Models}
Despite the sincere efforts made for explainable AI, there are still several challenges hindering its success and adoption. One of the key challenges is the interpretability of deep learning. In this regard, efforts are ongoing to develop explainable deep learning techniques or applications. To this aim, different visualization techniques are used to explain their reasoning steps, which is expected to make them explainable and trustworthy.

\subsubsection{Explainability and Adversarial AI}
As detailed earlier, explainability and adversarial AI has a direct connection. Explainability on the one side can guard against adversarial attacks by differentiating between a genuine sample and an adversary while on the other hand the information revealed by explainability techniques can be used both to generate more effective adversarial attacks on AI algorithms \cite{arrieta2020explainable}. One of the interesting directions of research on explainable AI is to analyze how effectively it can be used to guard against adversarial attacks. There are already ongoing efforts in this direction as detailed in Section \ref{explainable_adversarial_attacks}.

\subsection{AI Ethics}

Despite the significant progress AI ethics could make in a short period, many issues still remain open and various challenges still need to be addressed by future research. Below, we summarize the key points in this regard.

\begin{itemize}
 \item Due to its strongly interdisciplinary character and relatively young age, AI ethics suffers from serious conceptual ambiguity. Many of the key terms have fundamentally different and, sometimes even incompatible, meanings for different people. For example, key terms like agent, autonomy, and intelligence do not have the same meaning for moral philosophers and AI engineers. For engineers, cars or weapons will be ``autonomous'' when they can behave without direct human intervention. Moral philosophers, however, would use the term ``autonomous'' exclusively for an entity that can define its own laws or rules of behavior by itself \cite{powersethics}. To improve the AI moral discourse and make it more efficient, there is a dire need for future research that will enhance its conceptual clarity and standardize the primary and secondary meanings of its key terms.
 \item There is a need for exploring innovative ways to bridge the existing gaps between academic research and policymaking on one hand and between policymaking and the AI reality on the other hand. The questions raised and addressed by the academics are sometimes too abstract and theoretical to be of relevance for policymakers and those engaged in the AI business. Instead of broad philosophical questions like ``Will this contribute to human flourishing or put human species at risk?,'' policymakers are more interested in practical questions like ``Which harms should we expect if we are going to do this, and how to mitigate or minimize these harms?.'' Despite some good but still seemingly exceptional instances, various researchers also warn that there is hardly any touchable impact of ethics in general or policies and guidelines in particular on the reality of the AI industry. Most of the time, large companies are driven by economic logic and incentives rather than by value or principle-based ethics \cite{hagendorff2020ethics,korinek2019integrating}.
\item The moral discourse on AI systems is almost exclusively ``Western'' in nature. In other words, ethical deliberations and academic publications are published by institutions based in Western Europe and the United States and thus imbued with secular-oriented moral thought. With the expected growth of the AI industry and the adoption of its technologies by other communities worldwide, there is a need for diversifying and enriching the current AI moral discourse by incorporating insights from other cultural and religious traditions. Available research shows that people’s cultural norms do influence their understanding of what makes AI systems ethical \cite{awad2018moral}. Moreover, reports coming from Muslim-majority countries like Qatar, show that their interest in having AI technologies is espoused with a parallel interest in developing religio-culturally sensitive discourse and compliant policies, where also Arabic language processing will be a national priority \cite{QNAI}.

\end{itemize}

\section{Conclusions}
\label{conclusion}
In this paper, we have reviewed the key challenges in the successful deployment of AI in smart city applications. In particular, we focused on four key challenges namely security and robustness, interpretability, and ethical (data and algorithmic) challenges in the deployment of AI in human-centric applications. We particularly focused on the connection between these challenges and discussed how they are linked. Based on our analysis of the existing literature and experience in the domain, we also identified the current limitations and the pitfalls of existing solutions proposed for tackling these challenges. We also identify open research issues in the domain. We believe such a rigorous analysis of the domain will provide a baseline for future research. 

\section*{Acknowledgment}
This publication was made possible by NPRP grant \# [13S-0206-200273] from the Qatar National Research Fund (a member of Qatar Foundation). The statements made herein are solely the responsibility of the authors.

\bibliography{elsarticle-template}

\begin{thebibliography}{100}
\bibitem{huet2015server}
E.~Huet, Server and protect: Predictive policing firm {PredPol} promises to map
  crime before it happens, Forbes Magazine.
\bibitem{ackerman2019three}
E.~Ackerman, Three small stickers in intersection can cause {T}esla autopilot
  to swerve into wrong lane, IEEE Spectrum, April 1.
\end{thebibliography}

\end{document}